\documentclass[a4paper]{spie}  %>>> use this instead for A4 paper
\usepackage{amsmath,amsfonts,amssymb}
\usepackage{subcaption}
\usepackage{graphicx}
\usepackage[colorlinks=true, allcolors=blue]{hyperref}
\usepackage{siunitx}
\usepackage{acronym}

\usepackage{xcolor}
\usepackage{marginnote}

\DeclareMathOperator{\avg}{avg}

\newcommand{\pvec}[1]{\vec{#1}\mkern2mu\vphantom{#1}}
\newcommand{\svec}[3]{\vec{#1}{}^{\mkern7mu#2}_{#3}}

\acrodef{WST}{the Wide-field Spectroscopic Telescope}
\acrodef{Spec-S5}{the Spectroscopic Stage-5 Experiment}
\acrodef{MUST}{the MUltiplexed Survey Telescope}
\acrodef{SDSS}{the Sloan Digital Sky Survey}
\acrodef{PFS}{the Subaru Prime Focus Spectrograph}
\acrodef{4MOST}{the 4-metre Multi-Object Spectroscopic Telescope}
\acrodef{DESI}{the Dark Energy Spectroscopic Instrument}
\acrodef{MOS}{multi object spectroscopy}

\acrodef{QA}{quality assurance}
\acrodef{PWM}{pulse-width modulation}

\acrodef{DUT}{module under test}
\acrodef{PMMA}{acrylic}
\acrodef{SCARA}{Selective Compliance Assembly Robot Arm}

\acrodef{EPFL}{Swiss Federal Technology Institute of Lausanne}
\acrodef{SNSF}{Swiss National Science Foundation}

\acrodef{CMB}{Cosmic Microwave Background}
\acrodef{LSS}{large-scale structure}
\title{Characterizing robotic positioners under the influence of changing gravity vectors for future spectroscopic surveys }
\acrodef{BAO}{baryonic acoustic oscillations}
\acrodef{RSD}{redshift-space distortion}

\author[a,*]{Johannes Wüthrich}
\author[a]{Guandi Zhao}
\author[a]{Banan Yamani}
\author[a]{Léonard Lebrun}
\author[a]{Sean MacBride}
\author[a]{Andrin Fazan}
\author[a]{Felipe Andrade-Oliveira}
\author[a]{Marcelle Soares-Santos}
\affil[a]{Physik-Institut, Universität Zürich, Winterthurerstrasse 190, 8057 Zürich, Switzerland}

\authorinfo{Further author information: (Send correspondence to J.W.)\\
    \textsuperscript{*}J.W.: E-mail: johannes.wuethrich@physik.uzh.ch}

\begin{document}
\maketitle

\begin{abstract}
Future (Stage V) spectroscopic surveys intend to accurately map billions of galaxies.
To accomplish this goal, these surveys will employ highly multiplexed focal planes composed of robotic fiber positioners to accurately place individual optical fibers on targets of interest.
The ambitious science objectives place stringent requirements on the mechanical performance of these positioners.
Experience from previous surveys has shown that testing positioners under conditions closely resembling those on the telescope is of utmost importance during the prototyping and quality assurance phases of construction.
We present an automated telescope simulator test stand that characterizes the performance of these positioners at different orientations, reproducing the changing gravity vectors encountered during telescope operations.
The test stand aims to verify position stability down to \SI{1}{\micro\meter}, focus stability down to \SI{5}{\micro\meter}, as well as tilt variations lower than \SI{0.4}{\degree}.
We discuss the design of our setup, along with early characterization of image quality due to turbulence and the compensation of the enclosure deformation via calibration using fixed spots.
Finally, we present initial results of positioning stability tests using a prototype module built by Orbray Co., Ltd.
This test setup fulfills an important need for integrated testing of advanced focal plane prototypes under conditions similar to on-telescope conditions.
\end{abstract}


\section{Introduction}
\label{sec:intro}
\begin{figure}
    \begin{subfigure}[t]{0.48\textwidth}
        \centering
        \includegraphics[width=\textwidth]{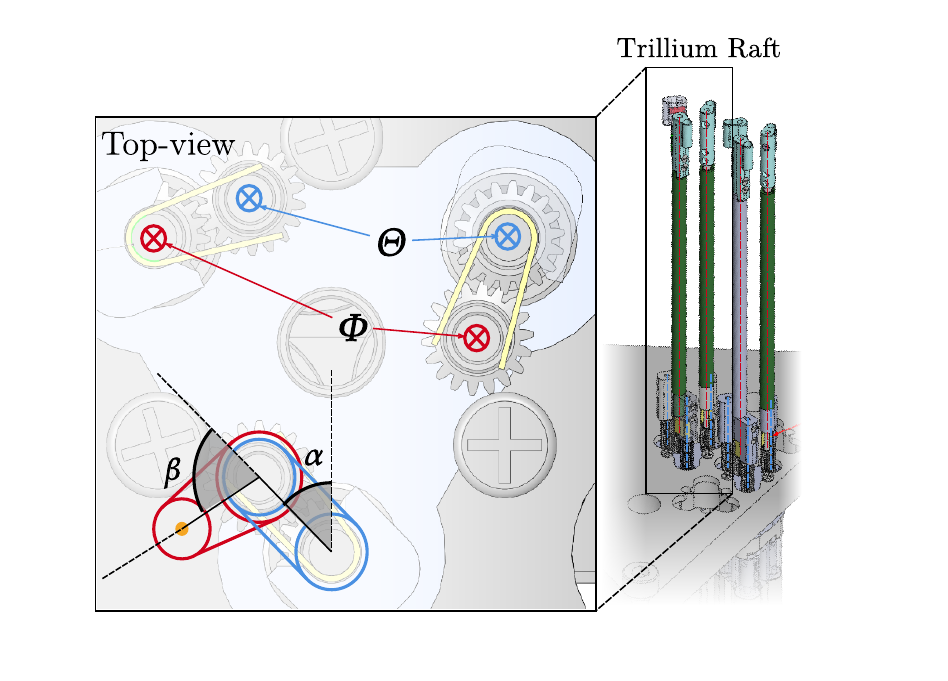}
        \caption{Illustration of the $\Theta$/$\Phi$ positioner kinematics with two angular degrees of freedom $\alpha$/$\beta$.}
        \label{fig:positioner_illustration_thetaphi}
    \end{subfigure}%
    \hfill
    \begin{subfigure}[t]{0.48\textwidth}
        \centering
        \includegraphics[width=0.95\textwidth]{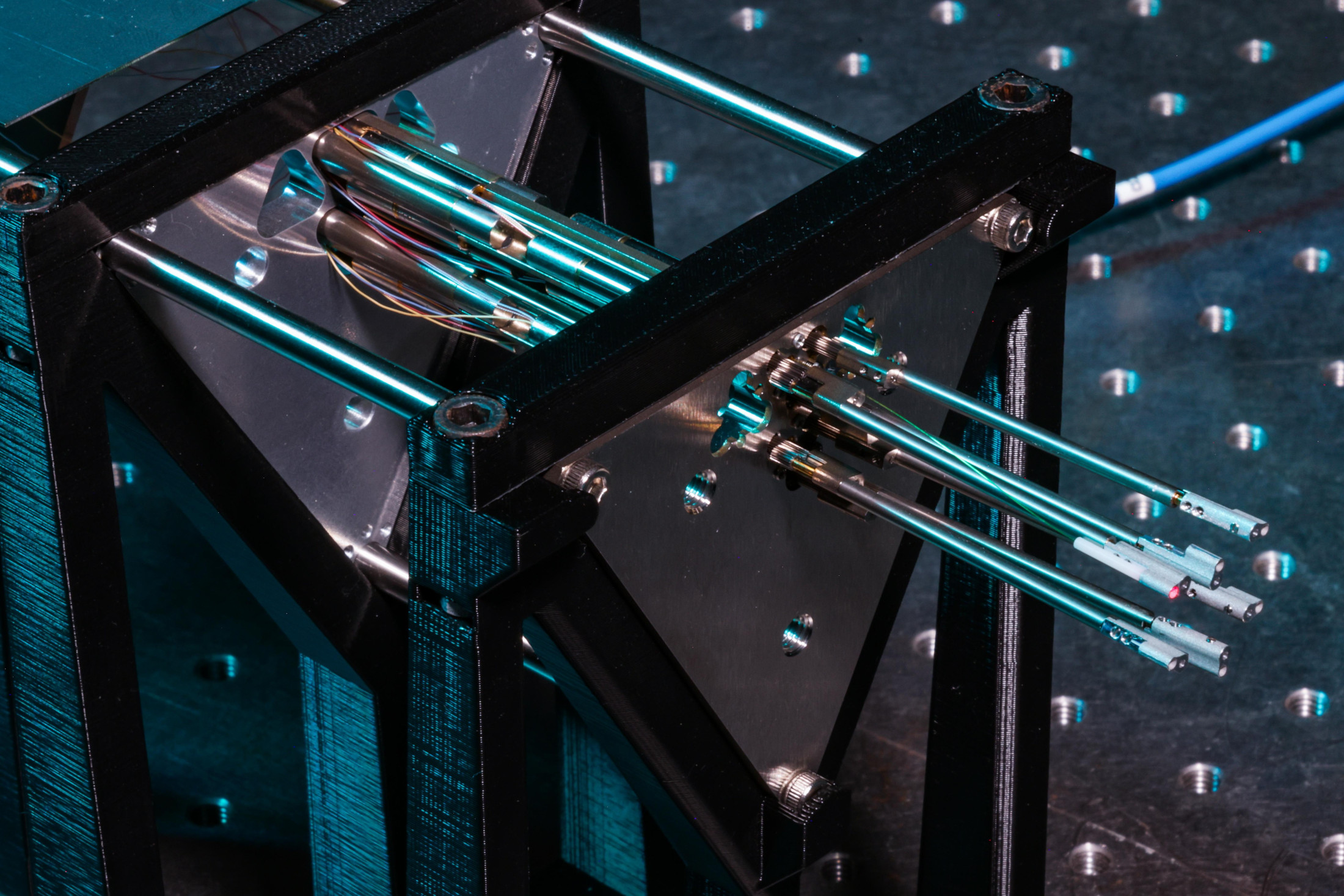}
        \caption{Image of a prototype multi-positioner module, showing an early prototype with only 6 out of 63 positioners populated -- Image by Alain Herzog / EPFL - CC-BY-SA 4.0.}
        % Image from: https://actu.epfl.ch/news/a-collaboration-to-push-the-boundaries-of-cosmos-m/
        \label{fig:positioner_illustration_module}
    \end{subfigure}
    \caption{Robotic fiber positioners: the $\Theta$/$\Phi$ kinematic concept (a) and a prototype module (b).}
    \label{fig:positioner_illustration}
\end{figure}

Spectroscopic measurements are the most precise method to measure the redshift of cosmological objects.
Fiber-fed \ac{MOS} instruments use optical fibers to collect the light from individual cosmological objects, enabling highly multiplexed precision measurements.
Early fiber-fed \ac{MOS} instruments, such as \ac{SDSS}, used plug plates to accurately position fibers on targets of interest.
This process involved the manufacturing of a dedicated aluminium plate for each exposure, with fibers manually placed into the pre-drilled holes of the plug-plate, a very labor-intensive process which practically limited the number of targets observed at the same time.
The introduction of robotic fiber positioners transformed the field and drastically increased the number of observed targets per observing night.
The fast and automatic reconfiguration of the focal plane, coupled with an increase in the fiber density, dramatically expanded the efficiency and scale of spectroscopic surveys.
The most widely used fiber positioner architecture is a two-stage \ac{SCARA} layout with two angular degrees of freedom, capable of placing the optical fiber within a circular patrol radius~\cite{silber_robotic_2023}, as illustrated in figure~\ref{fig:positioner_illustration_thetaphi}.
This type of positioner is in use in \ac{SDSS}-V and \ac{PFS}.
\Ac{DESI} has 5000 Theta/Phi positioners installed on its focal plane~\cite{silber_robotic_2023}, making it the most powerful \ac{MOS} instrument in use at the moment.
DESI recently completed its originally planned 5-year survey and mapped 47 million galaxies and quasars, thus creating the most complete 3D map of our universe to date~\cite{biron_desi_2026}.

The data from \ac{MOS} instruments have enabled a multitude of cosmology analyses, including measurements of the expansion history of the Universe via \ac{BAO}~\cite{alam_completed_2021,adame_desi_2025,desi_collaboration_desi_2025}, constraints on the growth rate of structure and tests of General Relativity using \ac{RSD}~\cite{alam_completed_2021,adame_desi_2025-1}, and measurements of the dark energy equation of state parameters~\cite{adame_desi_2025,desi_collaboration_desi_2025}.
Using these analyses, we have sharpened the constraints on fundamental cosmological parameters like the local rate of cosmic expansion $H_0$, the abundance of matter $\Omega_m$, and the equation of state of dynamical dark energy models $w_0$ and $w_a$.
In particular, \ac{BAO} measurements from \ac{DESI} yield $H_0 = (68.52 \pm 0.62)$\,km\,s$^{-1}$\,Mpc$^{-1}$~\cite{adame_desi_2025} which is consistent with measurements from \ac{CMB} anisotropies measured by the Planck experiment~\cite{aghanim_planck_2020-1}, but in tension with the local distance-ladder measurement from SH0ES~\cite{riess_comprehensive_2022-1}.
Future-generation spectroscopic surveys will improve these cosmological measurements by increasing the number of galaxies observed by approximately one order of magnitude while increasing the observed redshift range~\cite{besuner_spectroscopic_2025,zhao_multiplexed_2026}.
Next-generation spectroscopic surveys will significantly improve measurements of \ac{LSS} tracers, including \ac{BAO} and \ac{RSD}, by observing hundreds of millions of galaxy spectra, including large samples at high redshift ($z>2$).
These measurements will help probe the nature of dark energy and dark matter, and provide increasingly stringent tests of the $\Lambda$CDM model.
Fisher forecasts for \ac{MUST}, \ac{WST}, and other future surveys indicate that precision on \ac{BAO} parameters at the level of $\lesssim \SI{0.5}{\percent}$ may be achievable \cite{dassigniesd_cosmological_2023}.
The precise cosmography from the next generation of spectroscopic surveys will provide unprecedented constraints on the nature of dark energy, inflation models, neutrino physics, and dark matter, for example.

Future \ac{MOS} survey projects include \ac{Spec-S5}~\cite{besuner_spectroscopic_2025}, \ac{WST}~\cite{bacon_wst_2024,mainieri_wide-field_2024} and \ac{MUST}~\cite{zhao_multiplexed_2026}, which are in different stages of planning and design.
Each of these projects plans to employ between \SIrange{20000}{30000}{} positioners to allow for a higher survey multiplicity.
As the size of the focal plane is generally limited by the telescope optics, the increase in positioner number is achieved by reducing the size of each individual positioner and packing positioners more tightly together.
\ac{DESI} for example has a maximum patrol radius of ca.~\SI{6}{\milli\meter}~\cite{silber_robotic_2023}, while in the case of \ac{MUST} the planned patrol radius is only ca. \SI{3.6}{\milli\meter} for each positioner.
In existing surveys, the positioners were individually fabricated and tested prior to installation on the focal plane.
For future surveys (especially \ac{Spec-S5} and \ac{MUST}) the positioners are first integrated into (triangular) modules containing several dozen robots, which then undergo testing on a per-module basis.
\ac{MUST} for example plans on using ca.~\num{21000} robotic fiber positioners arranged in  modules with 63 positioners each, with a pitch of \SI{6.2}{\milli\meter}~\cite{rombach_investigations_2024}.
These modules will also contain the necessary drive electronics for all motors, and thus each module has a very minimalistic electrical and mechanical interface to the rest of the focal plane system.
This facilitates integration into the focal plane and also improves serviceability as individual modules can be easily replaced in the case of a positioner failure.
Figure~\ref{fig:positioner_illustration_module} shows an image of such a 63 positioner module.

In order to achieve the ambitious science goals of these future surveys, the performance of the positioners needs to be tested and verified extensively, both during the prototyping phase and as \ac{QA} during the production and integration phase.
Lessons learned from \ac{DESI} show that testing conditions should mirror the operational conditions on the telescope as closely as possible, especially with respect to the change in the gravity vector direction to which the positioners are exposed.
In the case of \ac{DESI} a significant fraction of positioners started showing degraded performance after being installed on the telescope focal plane.
The root cause of this issue was identified to be small fractures in a set of gears inside the positioner.
Positioners with these defects were unable to properly handle the upside-down orientation during the telescope operation.
As testing and verification of these positioners was carried out primarily with the positioner in a horizontal orientation on test benches, this issue was not identified prior to installation and operation on the telescope itself.
To date, approximately~\SI{15}{\percent} of the \ac{DESI} positioners show degraded performance~\cite{silber_robotic_2023}.

In this work, we present a dedicated test stand that allows us to test positioners for future surveys at different orientations relative to gravity.
The commissioning and initial characterization of this test-stand are described in detail in \cite{yamani_towards_2026}.
Our current focus lies in testing prototype positioner modules for the future \ac{MUST} survey and accordingly the test specifications are derived from the \ac{MUST} science specifications~\cite{zhao_multiplexed_2026,wei_must_2026}.
Overall we foresee four different types of tests to be carried out on the telescope simulator:
\begin{itemize}
  \item \textbf{XY-Stability}: Measurement of the (static) fiber placement stability while the telescope undergoes star tracking motions over a \SI{30}{\minute} period (target measurement resolution: \SI{<1}{\micro\meter}).
  \item \textbf{XY-Accuracy}: Various measurements of XY-repeatability, XY-linearity, datum repeatability and backlash (analogous to~\cite{rombach_investigations_2024}) at different telescope orientations (target measurement resolution: \SI{<5}{\micro\meter}).
  \item \textbf{Tilt}: Measurement of the tilt of the fiber and positioner axes relative to the optical axis, at different telescope orientations (target measurement resolution: \SI{<0.4}{\degree}).
  \item \textbf{Defocus}: Measurement of any displacement of the fiber tip along the optical axis, at different telescope orientations (target measurement resolution: \SI{<5}{\micro\meter}).
\end{itemize}
\begin{figure}
    \begin{subfigure}[t]{0.20\textwidth}
        \centering
        \includegraphics[width=\textwidth,clip,trim=35cm 0cm 28cm 0cm]{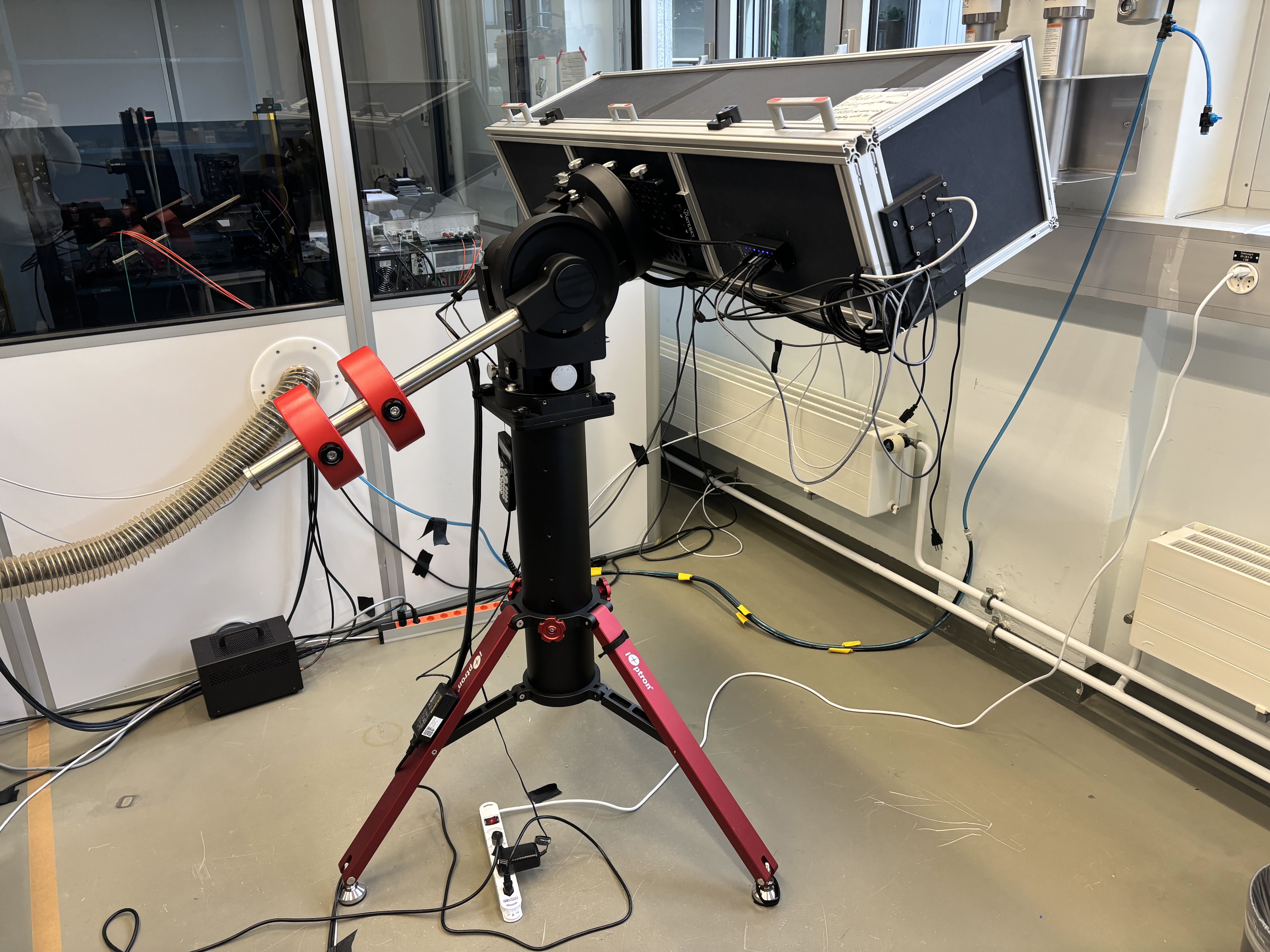}
        \caption{Dark box on the telescope mount.}
        \label{fig:testsetup_mount}
    \end{subfigure}%
    ~
    \begin{subfigure}[t]{0.78\textwidth}
        \centering
        \includegraphics[width=\textwidth,clip,trim=0 0.1cm 0.5cm 0]{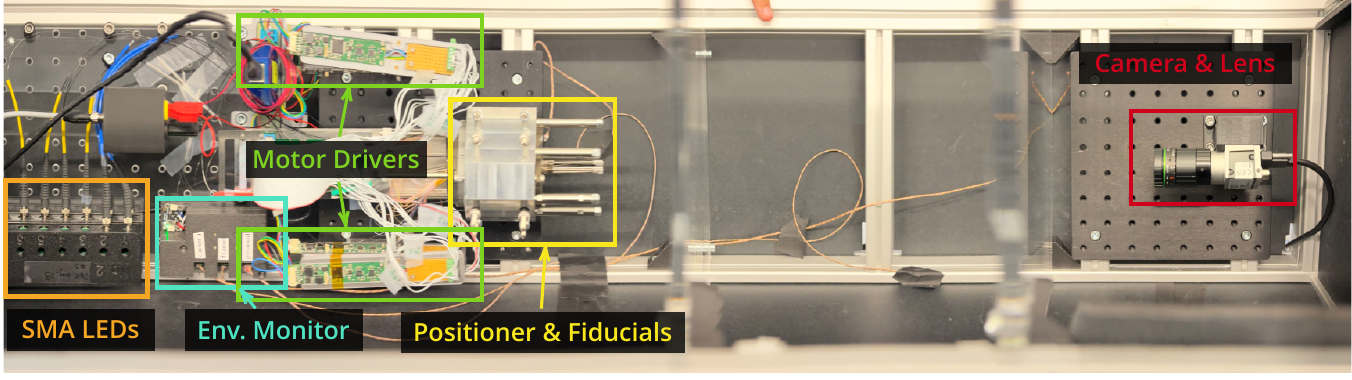}
        \caption{Annotated view of the components inside the dark box.}
        \label{fig:testsetup_inside}
    \end{subfigure}
    \caption{The telescope simulator test setup.}
    \label{fig:testsetup}
\end{figure}
The targeted measurement resolutions are based on the preliminary \ac{MUST} fiber positioner module specifications~\cite{wei_must_2026}.
The focus of this publication lies on general design and calibration of the test stand presented in section~\ref{sec:setup}, as well as on the measurements of the XY-Stability of a 6-positioner prototype module fabricated by Orbray Co.,~Ltd. which is further documented in section~\ref{sec:results}.
A setup to measure the tilt of the optical fiber at different telescope orientations is currently being developed following the approach discussed in~\cite{kronig_precision_2020,galal_prototyping_2025}.
Our work on measuring the change in the focus distance of the fiber at different telescope orientations is discussed separately in~\cite{lebrun_precision_2026}.
The general approach for XY-testing broadly follows previous work as published in~\cite{kronig_precision_2020,galal_prototyping_2025}, with optical fibers being back-illuminated, a computer vision camera placed in front of the positioner under test, and images taken by this camera being used to accurately detect and measure the position of the spot formed by the back-illuminated fiber.
In the case of XY-stability the goal is to verify that the fiber tip held by the robotic positioner stays in place during the duration of a single observation on the telescope.
For MUST such observations can take up to \SI{30}{\minute}, during which the telescope continuously tracks the night sky at an angular tracking rate of ca.~\SI{15}{\degree\per\hour}.
Thus the XY-Stability test setup needs to verify that a fiber tip does not drift by more than \SI{1}{\micro\meter} during a half-hour tracking movement.

\section{Measurement Setup}
\label{sec:setup}
An iOptron CEM120 equatorial mount is used as the foundation of the test setup using appropriate counter weights and supported by an iOptron Tri-Pier 360.
The equatorial mount is controlled via a serial protocol, enabling automatic measurements at different orientations.
A custom dark box built from aluminium extrusion profiles is mounted on the equatorial mount using a Losmandy adapter plate, ensuring the necessary stray light shielding for our metrology setup.
Figure~\ref{fig:testsetup_mount} shows an image of the mount and dark box in the lab.
Inside the dark box, the XY test setup is mounted on two small optical breadboards, with the metrology camera at one end and the positioner \ac{DUT} at the other end, together with auxiliary devices.
A Basler acA3800-14um camera is used together with a Fujinon HF3520 lens with a \SI{35}{\milli\meter} focal length.
The distance between the camera and the DUT is chosen so that the entire module (including additional fiducial spots) lies within the camera field of view.
Thorlabs FG105UCA multi-mode fibers are employed, with one end equipped with a \SI{1.25}{\milli\meter} ceramic ferrule and the other end with an SMA fiber connector.
Connectorization, ferrulization, polishing and fiber splicing are carried out in-house, and fiber ends are inspected using a fiber scope before installation in the setup.
The fibers are back-illuminated using custom modules containing eight Broadcom AFBR-1555ARZ SMA coupled LEDs, which are \ac{PWM} controlled using an Arduino Nano ESP32, allowing control of the light intensity in each fiber individually.
A second Arduino based module is installed inside the dark box, equipped with three temperature probes, an ambient temperature and humidity sensor, as well as an accelerometer which is used to provide feedback on the mount orientation during tests.
All devices are controlled via USB.
The \ac{DUT} is mounted using a custom-made aluminium holder providing the necessary rigidity for tests at different orientations.
The module is firmly clamped between aluminium pieces in this holder, and is therefore not expected to move relative to the holder, unless the module itself flexes.
A fiducial is mounted below the \ac{DUT} into the same aluminium holder.
The fiducial contains four \SI{10}{\micro\meter} diameter spots etched into a coated glass block and back-illuminated using an LED~\cite{baltay_desi_2019}.
Four additional fixed fibers are mounted surrounding the \ac{DUT}, acting as fixed reference points over a larger relative scale than the fiducial.
These fibers are mounted using stainless steel tubes screwed directly into the aluminium holder piece.
The steel tubes by design should show a maximum lateral flexing of \SI{0.2}{\micro\meter} under the effect of changing gravity directions.
The fiducial together with the fixed fibers are used to calibrate any systematic effects in the test setup, such as flexing of the dark box at different orientations, or thermal effects leading to mechanical distortion of the dark box.
A detailed description of the calibration is given in section~\ref{sec:calibration}.
\begin{figure}
    \begin{subfigure}[t]{0.5\textwidth}
        \centering
        \includegraphics[width=0.9\textwidth]{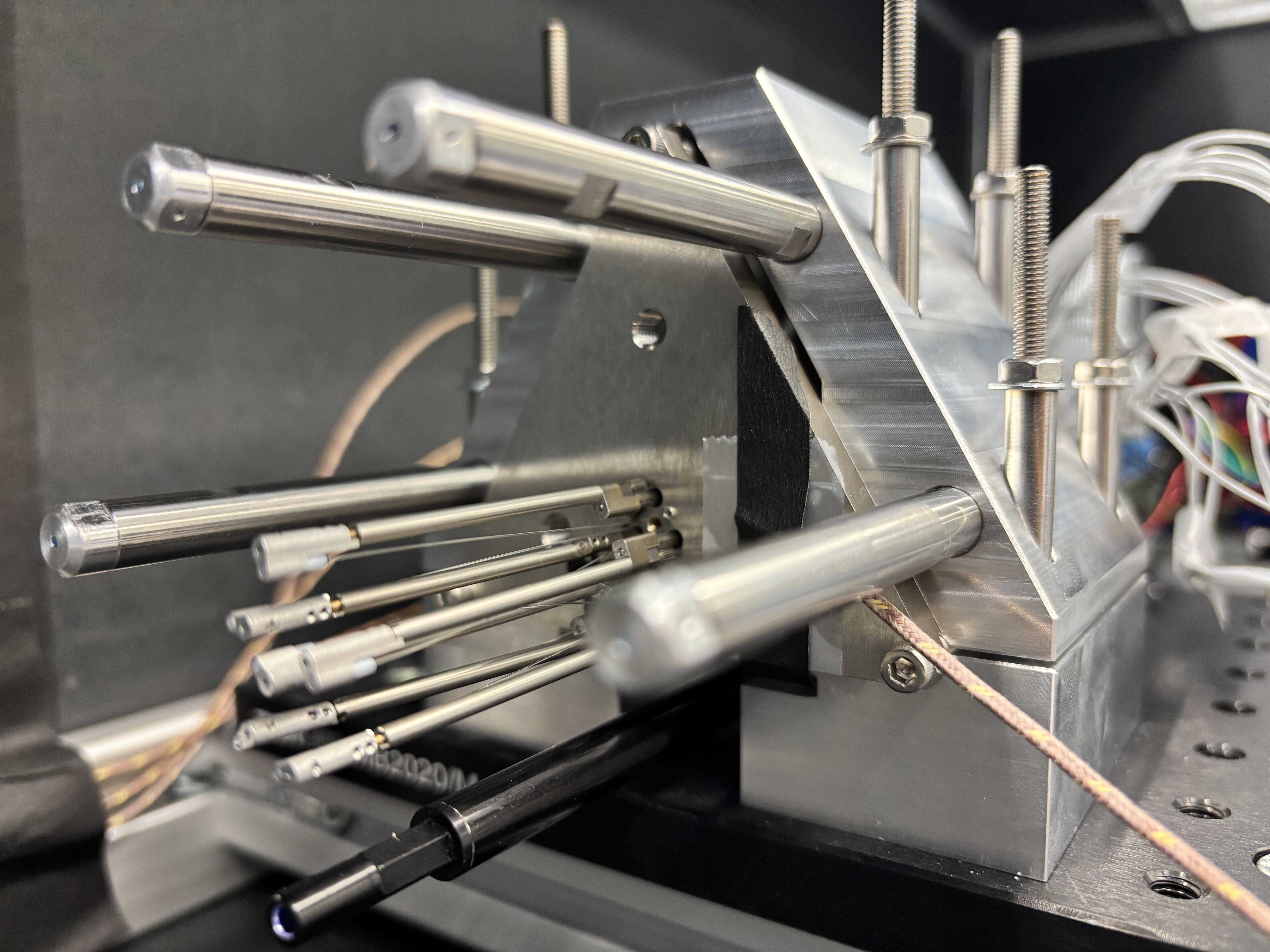}
        \caption{Image showing the Orbray-6 module installed.}
        \label{fig:testsetup_fixedspots_image}
    \end{subfigure}%
    ~
    \begin{subfigure}[t]{0.5\textwidth}
        \centering
        \includegraphics[width=0.9\textwidth]{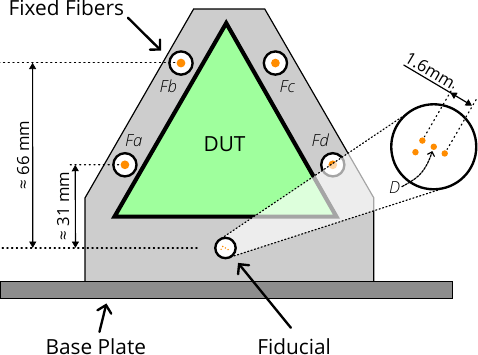}
        \caption{Field of view as seen by the camera.}
        \label{fig:testsetup_fixedspots_front_schematic}
    \end{subfigure}
    \caption{The mounted \ac{DUT}, surrounded by the fixed fibers and the fiducial.}
    \label{fig:testsetup_fixedspots}
\end{figure}
Figure~\ref{fig:testsetup_fixedspots_image} shows an image of the mounted \ac{DUT}, and a schematic view of the \ac{DUT} surrounded by the fiducial and the fixed fibers as seen by the camera is shown in figure~\ref{fig:testsetup_fixedspots}.

The initial testing campaign focuses on measurements with the telescope mount at different altitude angles, keeping the azimuth angle constant for most measurements.
We define three main orientations as $\theta = \SI{-90}{\degree}$ when the dark box is in a vertical orientation with the camera below and the positioner pointing towards the ground, $\theta = \SI{90}{\degree}$ the opposite case with the positioners point towards the sky, and $\theta = \SI{0}{\degree}$ when the dark box is in a horizontal configuration.
All measurements are strictly carried out within the range of $\theta \in \SIrange{-90}{89}{\degree}$ due to technical limitations of the mount.
During normal operation on the \ac{MUST} survey telescope, the positioners will only experience altitude angles between \SIrange{0}{90}{\degree}.

\subsection{Setup Optimizations}
For any type of measurement carried out on this test-stand, the fundamental test sequence is always the same:
\begin{enumerate}
    \item Enable the fiber and fiducial back-illumination.
    \item Bring the system into a specified state (altitude $\theta$, azimuth $\phi$, positioner configuration $\alpha, \beta$, etc.).%
      \footnote{During certain tests, namely the XY-stability tests, the telescope mount is set to execute a slow but continuous star tracking motion, and thus images/measurements are taken while the mount slowly moves.}
    \item Take an image of the back-illuminated fiber and fiducial spots.
    \item Run a spot detection algorithm and fit each spot with a 2D Gaussian model.
    \item Store the fitted spot positions for later analysis.
    \item Go back to step 2 and the next system state.
\end{enumerate}
In order for the Gaussian spot fitting to produce good results, there are a number of guidelines to be followed: All light spots should have roughly the same intensity, which is achieved by individually tuning the light output for each fiber and fiducial spot via the \ac{PWM} controlled back-illumination.
No saturation should occur in the recorded image, which is achieved by adjusting the camera exposure time so that the maximum pixel intensity is ca.~\SI{90}{\percent} of the camera full scale.
The spots themselves should ideally have a true 2D Gaussian profile.
To ensure this we polish both ends of the fibers and couple the fibers to the back-illumination LEDs using SMA connectors.
In addition, we limit our aperture to an $f$-number of 8, as low $f$-numbers lead to a noticeable coma, while high $f$-numbers lead to Airy disk patterns.%
\footnote{While a good spot shape is important for XY tests, it is of much higher importance for defocus tests, as discussed in~\cite{lebrun_precision_2026}.}

\subsubsection{Shot-Noise Reduction}
\begin{figure}
    \centering
    \includegraphics[width=0.533\textwidth]{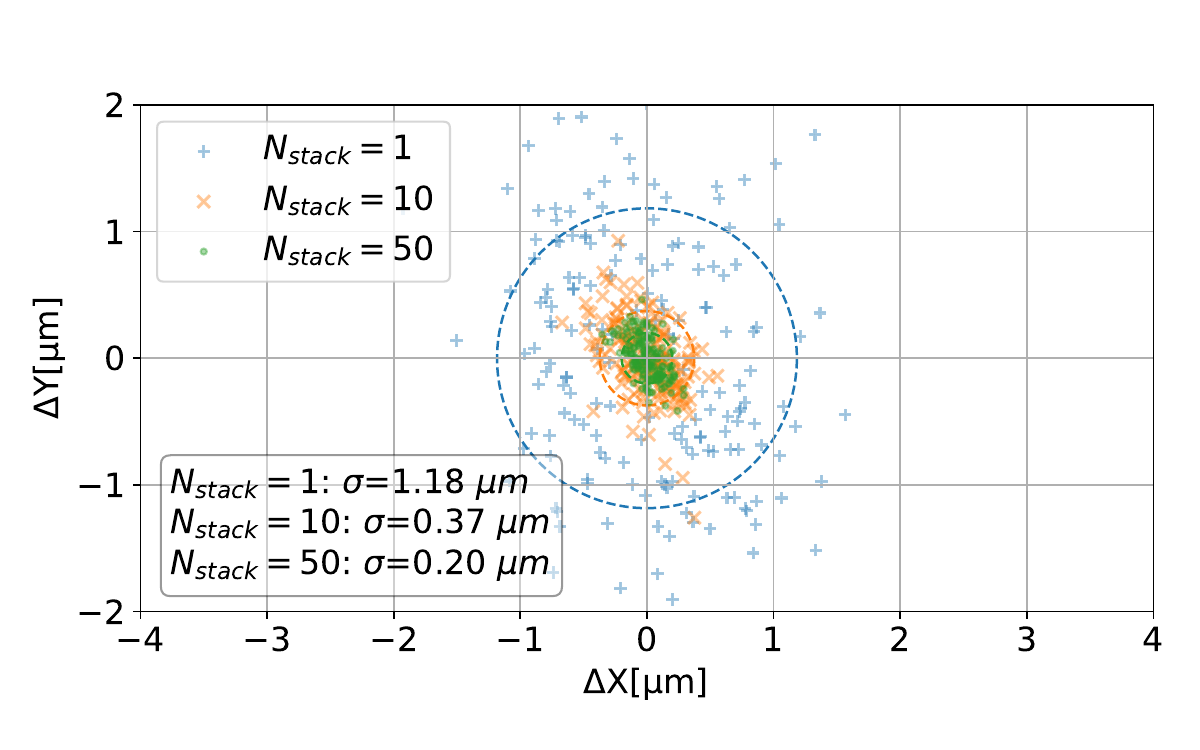}
    \caption{The effect of image stacking on spot stability after calibration. Fixed fiber $Fa$ is calibrated based on the position of other fixed fibers and measured with $N_{stack}= 1,~10 ~ \text{and} ~ 50$. The circles indicate the 1$\sigma$ of fiber position scatter.}
    \label{fig:shot_noise_reduction}
\end{figure}

The statistical uncertainty of the amplitude $\sigma_{A_{ij}}$ recorded in each pixel of the camera image is the dominant statistical noise source in our measurements.
This amplitude uncertainty directly translates into an uncertainty of the fitted spot positions $\sigma_{\vec{\mu}_k}$~\cite{winick_cramerrao_1986}.
We assume the amplitude uncertainty to be dominated by a Poissonian shot-noise process in the camera, and thus expect $\sigma_{A_{ij}} \propto \frac{1}{\sqrt{N_{stack}}}$ when averaging $N_{stack}$ consecutive camera frames (image stacking).
Accordingly we also expect $\sigma_{\vec{\mu}_k} \propto \frac{1}{\sqrt{N_{stack}}}$ when using image stacking.
Figure~\ref{fig:shot_noise_reduction} shows the effect on the position uncertainty for three different values: $N_{stack}= 1,~10 ~ \text{and} ~ 50$, by measuring the calibrated position of fiducial fixed fiber $Fa$, with calibration details discussed later in section~\ref{sec:calibration}.
The radius uncertainty decreases from $\SI{1.2}{\micro\meter}$ at $N_{stack}=1$ to $\SI{0.37}{\micro\meter}$ at $N_{stack}=10$ by a factor of $3.17\approx\sqrt{10}$, consistent with the position noise being dominated by Poissonian noise.
When increasing to $N_{stack}=50$, the position uncertainty reaches $\sim\SI{0.2}{\micro\meter}$, suppressed further by a factor of $1.87<\sqrt{5}$, which indicates hitting a noise floor that is not sourced from shot noise.
For our XY measurements we generally take $N_{stack}=50$ for optimal noise suppression.
The main trade-off from stacking a large number of images is the increase in acquisition time, due to the limited frame-rate of the camera of ca.~\SI{14}{\hertz}.
In addition, using $N_{stack}=50$ stacking acts as an intrinsic low-pass filter, smoothing out any measured fluctuations with time scales $\lesssim\SI{3.5}{\second}$.

\subsubsection{Thermal Turbulence Mitigation}
\begin{figure}[b]
    \begin{subfigure}[t]{0.22\textwidth}
        \centering
        \includegraphics[width=0.95\textwidth]{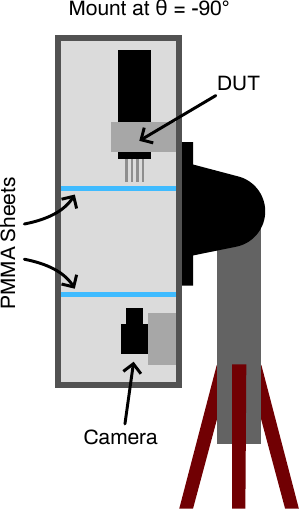}
        \caption{Sketch of the \ac{PMMA} sheets installed in the dark box.}
        \label{fig:turbulences_sketch}
    \end{subfigure}%
    \hfill
    \begin{subfigure}[t]{0.72\textwidth}
        \centering
        \includegraphics[width=\textwidth,clip,trim=0 3.5cm 0 3.5cm]{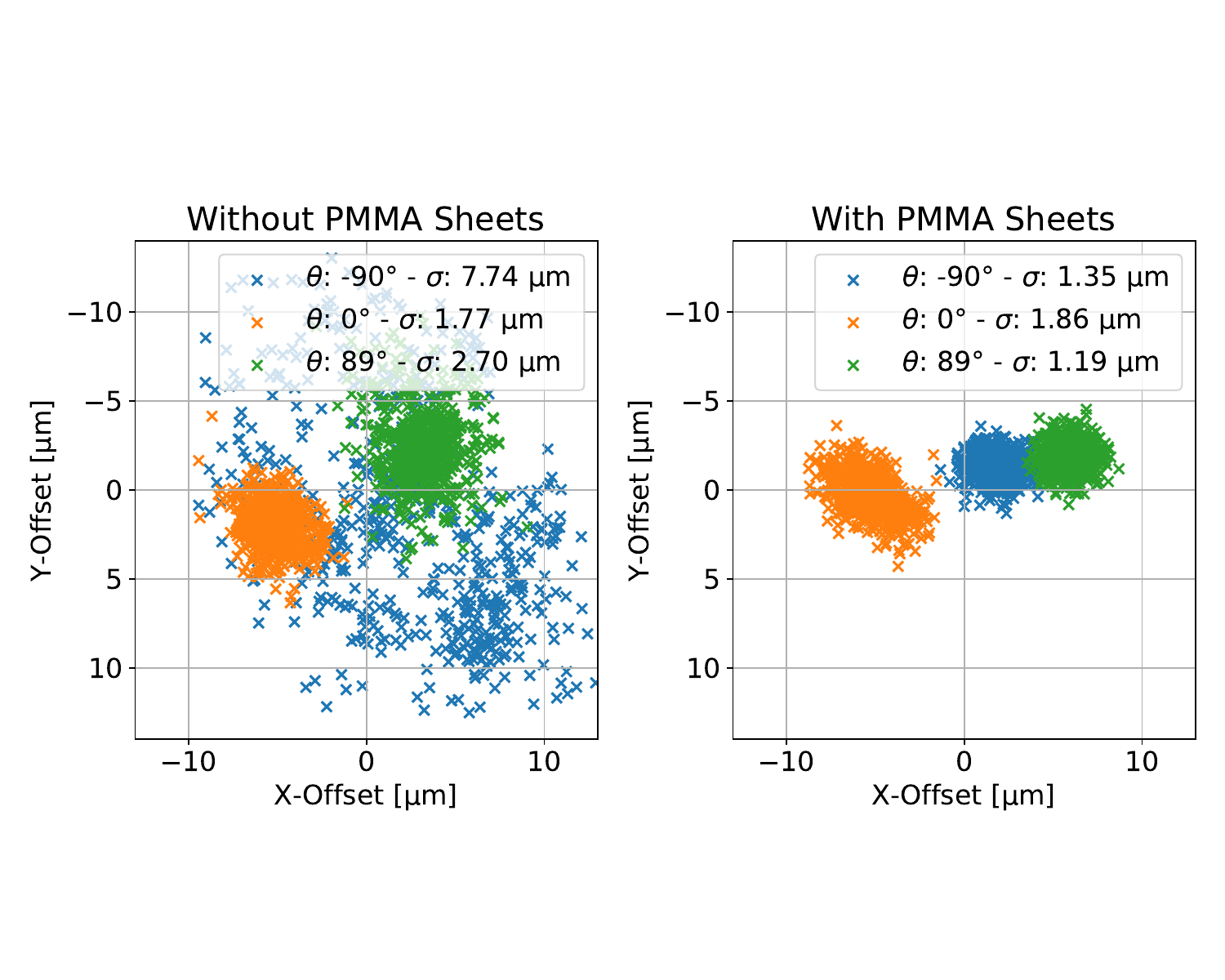}
        \caption{Stability before and after installation of \ac{PMMA} sheets. Note: These measurements were taken without image stacking and less light-efficient fibers compared to the final measurements.}
        \label{fig:turbulences_results}
    \end{subfigure}
    \caption{%
      Investigation of extra measurement uncertainty at \SI{-90}{\degree} caused by air turbulences.
      Installing the transparent \ac{PMMA} sheets leads to a strong improvement of the measured spot precision at $\theta = \SI{-90}{\degree}$, as well as a smaller improvement at \SI{89}{\degree}.
    }
    \label{fig:turbulences}
\end{figure}

Initial trial measurements had shown that the overall precision deteriorated with the mount approaching $\theta \rightarrow \SI{-90}{\degree}$ (thus having the positioner pointing down with the camera below).
Investigations indicated that the extra measurement uncertainty could be caused by air turbulences within the dark box.
With the camera located at the bottom of the system, it would act as a heat source, generating convective air flow that rises through the space between the camera and the \ac{DUT} and distorts the recorded image.
To verify this hypothesis two transparent \ac{PMMA} sheets were installed within the dark box between the camera and the DUT, effectively dividing the overall air volume into three sub-volumes, with no direct air-flow between the sub-volumes, as indicated in figure~\ref{fig:turbulences_sketch}.
Figure~\ref{fig:turbulences_results} compares the measurement stability before and after installation of the \ac{PMMA} screens, showing a clear improvement.
These results are a strong indication that thermal air turbulences can have an important effect on the data, and all subsequent measurements are taken with the \ac{PMMA} screens installed.

\subsection{Acquisition Calibration}
\label{sec:calibration}

In an ideal and static test setup, with the camera sensor plane exactly parallel to the object plane, the mapping from object space $\vec{r}_{ideal}$ (fiber tip) to pixel space $\vec{P}_{ideal}$ (spot location on the image) is given by a simple projection
\begin{equation}
    \vec{P}_{ideal} = s \cdot \left(\vec{r}_{ideal} - \vec{r}_{offset} \right)
\end{equation}
with $s$ a constant scale factor and $\vec{r}_{offset}$ an arbitrary constant offset, both determined by calibration.
In our test setup this simplistic ideal mapping does not hold for multiple reasons: the dark box does not have infinite rigidity and thus will flex when exposed to different gravity orientations leading to a relative shift of the image and spot planes.
In addition, temperature effects can further lead to distortions of the dark box as well as slightly changing the camera lens focal distance.
We develop a more complex calibration transformation between object and pixel spaces, assuming that all measured points lie on a single plane in the object space $\vec{r} = (x, y, 0)$ and that the object plane does not experience any mechanical distortion.
We allow the object plane to rotate freely in 3D space around an arbitrary rotation center relative to a reference frame whose $x'$ and $y'$ axis are aligned with the respective camera sensor axis, giving us
\begin{equation}
  \pvec{r}' = R_z(\gamma_z) R_y(\gamma_y) R_x(\gamma_x) \left(\vec{r} - \vec{r}_{rotation}\right) + \vec{r}_{rotation} = \begin{pmatrix}
      r'_x \\ r'_y \\ r'_z
    \end{pmatrix}. \label{eq:rotation}
\end{equation}
where $R_x, R_y, R_z$ are 3D rotation matrices.
By design, the object plane is nearly parallel with the camera plane, and thus we expect $\gamma_x$ and $\gamma_y$ to be close to $0$.
Any small rotation around $x$ and $y$ can be interpreted as a small out of plane tilt of the object plane, for example due to deformation of the dark box when changing the gravity direction.
Accordingly, one would also expect the center of rotation $\vec{r}_{rotation}$ to be close to the center of the object plane.
The transformed coordinates $\pvec{r}'$ are then mapped to pixel coordinates via a standard camera (pin-hole) projection as
\begin{equation}
  \vec{P} = A \begin{pmatrix}
    \dfrac{r'_x}{D + r'_z} \\
    \dfrac{r'_y}{D + r'_z}
    \label{eq:projection}
  \end{pmatrix}
\end{equation}
where $A$ is a proportionality factor which absorbs the angular field of view as well as the number of pixels of the camera, and $D$ is the optical distance between the object plane and the camera image plane.
Combining \eqref{eq:rotation} and \eqref{eq:projection}, and simplifying, we obtain
\begin{equation}
  \vec{P} = \frac{A}{D + y \cos(\gamma_y) \sin(\gamma_x) - x \sin(\gamma_y)}
  % Fully multiplied version
  % \begin{pmatrix}
  %   x \cos(\gamma_y) \cos(\gamma_z) + y \left( \cos(\gamma_z) \sin(\gamma_x) \sin(\gamma_y) - \cos(\gamma_x) \sin(\gamma_z) \right) + x_0 \\
  %   x \cos(\gamma_y) \sin(\gamma_z) + y \left( \cos(\gamma_z) \cos(\gamma_x) + \sin(\gamma_x) \sin(\gamma_y) \sin(\gamma_z) \right) + y_0
  % \end{pmatrix}
  \left(
    \begin{pmatrix}
      \cos(\gamma_z) & -\sin(\gamma_z) \\
      \sin(\gamma_z) & \cos(\gamma_z)
    \end{pmatrix}
    \begin{pmatrix}
      x \cos(\gamma_y) + y \sin(\gamma_y) \sin(\gamma_x) \\
      y \cos(\gamma_x)
    \end{pmatrix}
    +
    \begin{pmatrix}
      x_0 \\
      y_0
    \end{pmatrix}
  \right)
  \label{eq:rotation_calibration}
\end{equation}
where $x_0$ and $y_0$ absorb any terms which do not depend on $x$ and $y$.
This leaves us with a calibration model with 7 free parameters that we assume can change with time: $A(t)$, $D(t)$, $\gamma_x(t)$, $\gamma_y(t)$, $\gamma_z(t)$, $x_0(t)$ and $y_0(t)$.
The time $t$ here is a proxy for the underlying physical effect leading to a change, such as any change in temperature, or the change in the gravity direction (and thus the dark box flexing) depending on the mount angle $\theta$.
The calibration parameters are calculated for each acquired image, making use of the fixed fiber spots~$Fa$, $Fb$, $Fc$, $Fd$ and the spot~$D$ at the center of the fiducial (see figure~\ref{fig:testsetup_fixedspots_front_schematic}).
In order to employ minimization to calculate the calibration parameters, the position of these five spots in the object domain needs to be known.
Spot~$D$ is placed at the origin.
The locations of the other four spots was measured by rotating the mount to $\theta = \SI{-90}{\degree}$ and acquiring 120 images at \SI{30}{\second} intervals.
Equation~\eqref{eq:rotation_calibration} and the known locations of the four fiducial spots were used as a calibration during this preparatory run.
The averaged positions from the 120 images are used as the ground truth for $\vec{r}_{Fa}$, $\vec{r}_{Fb}$, $\vec{r}_{Fc}$ and $\vec{r}_{Fd}$ to be used for calibration going forward, with approximate locations indicated in figure~\ref{fig:testsetup_fixedspots_front_schematic}.
It should be noted, given that the goal of this setup is to measure relative changes between different telescope orientations, and not to measure absolute positions, that the precise values of $\vec{r}_{Fa}, \ldots, \vec{r}_{Fd}$ are not of absolute importance.
A (high) uncertainty on the true value of $\vec{r}_{Fa}$ would lead to at maximum ca.~\SI{0.3}{\percent} systematic error on the absolute measured positions $\vec{r}$.
But, importantly, as the \textit{true} values of $\vec{r}_{Fa}, \ldots, \vec{r}_{Fd}$ are fixed, this systematic error is constant across all measurements, and thus does not influence any comparative or relative measurements.

\subsubsection{Impact of Fibers offset from the Ideal Object Plane}
\label{sec:calibration_outofplane}
So far an underlying assumption has been that all image spots lie on a single physical plane.
In order to investigate the effect of small deviations from this assumption, we calculate the result of the rotation transformation~\eqref{eq:rotation} for a non-ideal spot located at $\vec{r}_{\Delta{}z} = (x, y, \Delta{}z)$, where $\Delta{}z$ represents the offset from the ideal object plane.
Propagating this shift through the rotational transform we obtain
\begin{equation}
  \Delta\pvec{r}' = \pvec{r}'_{\Delta{}z} - \pvec{r}' = \Delta{}z \begin{pmatrix}
       \cos(\gamma_z) \sin(\gamma_y) \cos(\gamma_x) + \sin(\gamma_z) \sin(\gamma_x) \\
       \sin(\gamma_z) \sin(\gamma_y) \cos(\gamma_x) - \cos(\gamma_z) \sin(\gamma_x) \\
       \cos(\gamma_y) \cos(\gamma_x)
    \end{pmatrix} = \begin{pmatrix}
       \Delta{}r'_x \\
       \Delta{}r'_y \\
       \Delta{}r'_z
    \end{pmatrix}\label{eq:dr_error}
    % = \begin{pmatrix}
    %   \cos(\gamma_z) & -\sin(\gamma_z) & 0 \\
    %   \sin(\gamma_z) & \cos(\gamma_z) & 0 \\
    %   0 & 0 & 1
    % \end{pmatrix}
    % \begin{pmatrix}
    %   \cos(\gamma_x) \sin(\gamma_y) \\
    %   -\sin(\gamma_x) \\
    %   \cos(\gamma_y) \cos(\gamma_x)
    % \end{pmatrix}
\end{equation}
with $\Delta{}r'_x, \Delta{}r'_y$ representing out-of-plane tilt-coupling terms and $\Delta{}r'_z$, a radial scaling term.
We analyse two different scenarios separately: $\Delta{}z$ constant for a given spot, and $\Delta{}z$ varying with time.
For $\Delta{}z$ constant, the $\Delta\pvec{r}'_z$ component leads to a relative error on the optical distance factor of the order of
\begin{equation*}
  \frac{D + \Delta{}z}{D} - 1 \approx \SI{0.29}{\percent} ~ \text{for} ~ D = \SI{35}{\centi\meter} ~ \text{and} ~ \Delta{}z = \SI{1}{\milli\meter}.
\end{equation*}
with $D$ taken from the setup calibration (see figure~\ref{fig:calibration_all_parameters}) and $\Delta{}z = \SI{1}{\milli\meter}$ a worst case estimate.
This value only changes by ca.~\SI{5e-4}{\percent} for angles $\SI{0}{\degree} \leq \gamma_x, \gamma_y \leq \SI{2}{\degree}$ leading to a sub-\si{\micro\meter} error on radial distances $\leq \SI{10}{\centi\meter}$.
This estimate is very conservative, as from experience the angles $\gamma_x$ and $\gamma_y$ at most change on the order of \SI{0.2}{\degree} over the entire altitude range $\SI{-90}{\degree} \leq \theta \leq \SI{90}{\degree}$

% Removed, as it does not really contribute much
% \begin{figure}
%     \centering
%     \includegraphics[width=0.5\textwidth]{figures/20260615_DzErrorPlots/DzConstant_XY.pdf}
%     \caption{%
%       Error in $x'$ and $y'$ in function of the $\Delta{}z$ shift and the tilt angles $\gamma_x, \gamma_y$.
%     }
%     \label{fig:r_error}
% \end{figure}
A given $\Delta{}z$ offset leads to an apparent $x', y'$ shift on the order of
\begin{equation}
  r'_{error} = \sqrt{(\Delta{}r'_x)^2 + (\Delta{}r'_y)^2} = \Delta{}z \sqrt{\sin^2(\gamma_x) + \cos^2(\gamma_x) \sin^2(\gamma_y)} < \sqrt{2}\Delta{}z |\sin(\max(\gamma_x, \gamma_y))|\label{eq:rerror}
\end{equation}
% Figure~\ref{fig:r_error} shows the resulting error for $\Delta{}z \in \SIrange{0}{1}{\milli\meter}$ and $\max(\gamma_x, \gamma_y) \in \SIrange{0}{1}{\degree}$.
For a comparative measurement, only the variation of $r'_{error}$ as $\gamma_x$ and $\gamma_y$ change with the mount orientation contributes a spurious drift.
This drift is estimated as:
\begin{equation}
  \delta r'_{error} < \sqrt{2}\Delta{}z |\max(\delta \gamma_x, \delta \gamma_y)|.
\end{equation}
With an estimate of $\Delta z = \SI{1}{\milli\meter}$ from installation of the \ac{DUT}, and a maximum change of $\delta \gamma_{max} <\SI{0.05}{\degree}$ in calibration angles $\gamma_x,\gamma_y$ over a \SI{30}{\minute} window, this induced drift is $\delta r'_{error} < \SI{1.2}{\micro\meter}$.
The values chosen for this estimate are very conservative, especially $\delta \gamma_{max}$ should be much lower for most measurements.

\begin{table}[t]
  \centering
  \caption{%
    Summary of the apparent spot-position errors caused by an out-of-plane offset $\Delta{}z$, separated by mechanism (radial scaling vs. out-of-plane tilt coupling) and by whether the offset is constant or varies in time, $\delta(\Delta{}z)$.
  }
  \label{tab:dz_summary}
  \begin{tabular}{lcl}
    \hline
    Scenario & Apparent shift & Order of magnitude \\
    \hline
    Constant $\Delta{}z$ -- radial scaling & $(r/D)\,\Delta{}z$ & sub-\si{\micro\meter} relative drift \\
    Constant $\Delta{}z$ -- tilt coupling & $\sqrt{2}\,\Delta{}z \delta\gamma$ & $< \SI{1.2}{\micro\meter}$ per exposure (conservative) \\
    Time-varying $\Delta{}z$ -- radial scaling & $(r/D)\,\delta(\Delta{}z)$ & \SI{1}{\micro\meter} per $\delta(\Delta{}z) = \SI{9}{\micro\meter}$ at $r = \SI{40}{\milli\meter}$ \\
    Time-varying $\Delta{}z$ -- tilt coupling & $\sqrt{2}\delta(\Delta{}z) \sin\gamma_\text{eff}$ & \SI{1}{\micro\meter} per $\delta(\Delta{}z) = \SI{20}{\micro\meter}$ at $\gamma_\text{eff} \approx \SI{2.0}{\degree}$ \\
    \hline
  \end{tabular}
\end{table}

A changing $\Delta z(t)$ scenario can happen if the \ac{DUT} positioner tip experiences out-of-plane motion.
In this case, the main contribution of $\Delta\pvec{r}'$ is the coupling of the $\Delta{}r'_z$ component into the projection~\eqref{eq:projection}.
Since the calibration assumes $z = 0$, a spot with offset $\Delta{}z$ and radial distance $r$ from the optical axis is recovered at an object position scaled by $r' \approx r \frac{D}{D + \Delta{}z}$, leading to a time dependent variation of
\begin{equation}
  \Delta{}r'(t) \lesssim -\frac{r}{D}\,\delta(\Delta{}z(t)). \label{eq:dz_radial}
\end{equation}
With a conservative estimate of $r \lesssim \SI{4}{\centi\meter}$, and nominal $D$ value, a $\delta(\Delta{}z)\approx\SI{9}{\micro\meter}$ motion translates to a lateral drift of $\Delta{}r' \lesssim \SI{1}{\micro\meter}$, with fiber spots closer to the optical axis less affected.

An out-of-plane motion also contributes to the out-of-plane tilt-coupling term of~\eqref{eq:rerror}, contributing
\begin{equation}
  \delta r'_{error}(t) \lesssim \sqrt{2}\,|\sin(\max(\gamma_x, \gamma_y))|\;\delta\left(\Delta{}z(t)\right). \label{eq:dz_tilt}
\end{equation}
This contribution is proportional to the absolute planar angle between camera perspective and fibers.
With a conservative upper limit of $\max(\gamma_x, \gamma_y) = \SI{2}{\degree}$, this translates a $\delta(\Delta z)\approx\SI{20}{\micro\meter}$ axial shift into a \SI{1}{\micro\meter} apparent drift on the XY plane.

A summary of the discussed $\Delta{}z$ effects is given in table~\ref{tab:dz_summary}.
An important limitation of an XY-only test setup is that it cannot distinguish between a physical lateral drift and a $\delta(\Delta{}z)$ drift.
In order to disentangle the apparent lateral shift due to $\delta(\Delta{}z)$ we would need to be able to measure the axial shift with a precision of $\Delta z_{error}\lesssim \SI{10}{\micro\meter}$.
In a companion paper~\cite{lebrun_precision_2026} we discuss a solution for relative $z$ distance measurements using depth from defocus with a precision of $\sim \SI{5}{\micro\meter}$.
Combining the XY setup with the depth-from-defocus test setup might allow to better constrain XY motion in the future.

\subsection{Measurement Setup Stability}
\label{sec:stability}
\begin{figure}
    \begin{subfigure}[t]{0.62\textwidth}
        \centering
        \includegraphics[width=\textwidth]{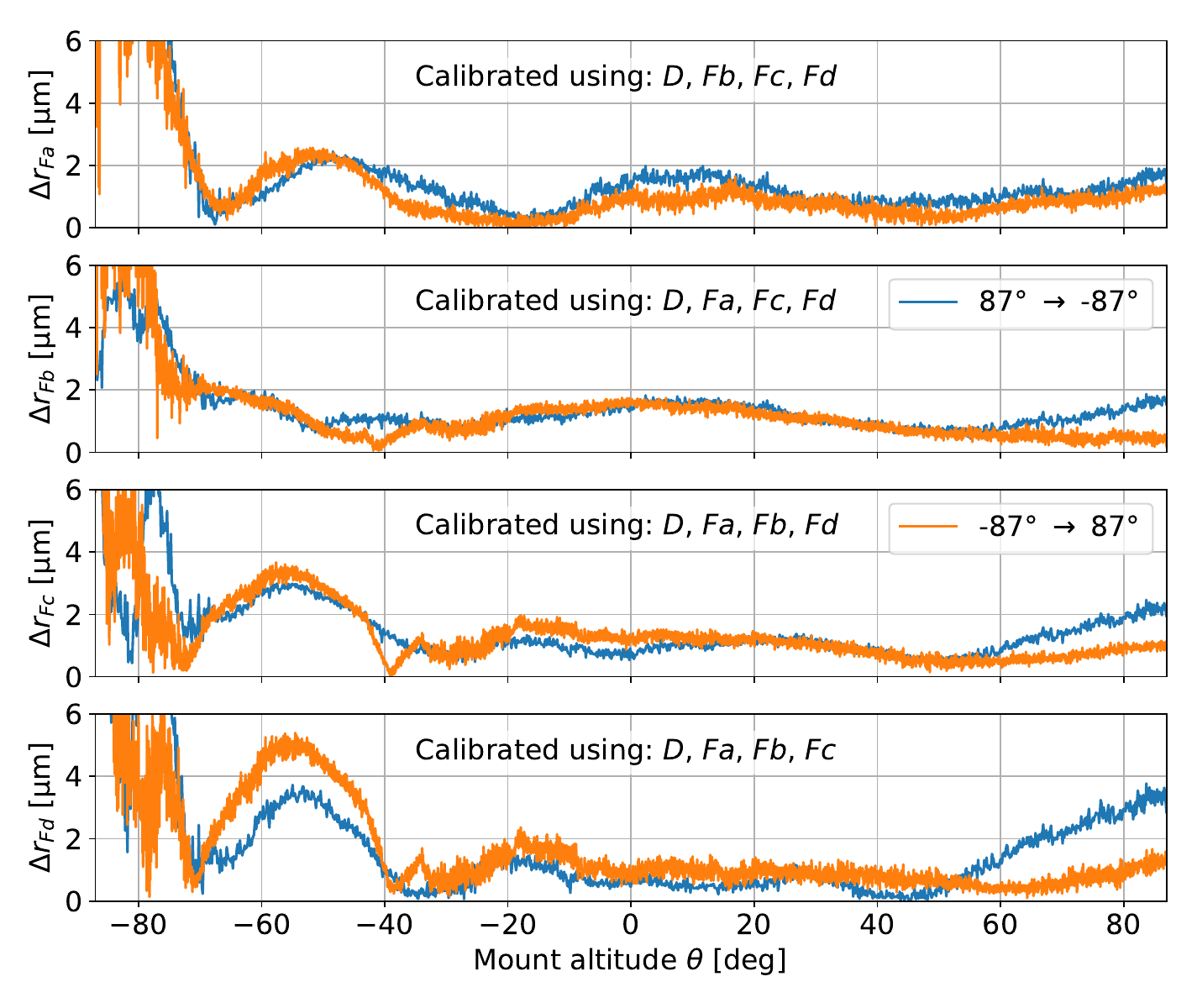}
        \caption{%
          Absolute deviation of the fixed test spot position relative to the average spot location over the entire range \SIrange{-87}{87}{\degree} for all four fixed fibers.
          The measured deviation does not (substantially) depend on the mount tracking direction.
        }
        \label{fig:calibration_test_Full}
    \end{subfigure}%
    \hfill
    \begin{subfigure}[t]{0.36\textwidth}
        \centering
        \includegraphics[width=\textwidth]{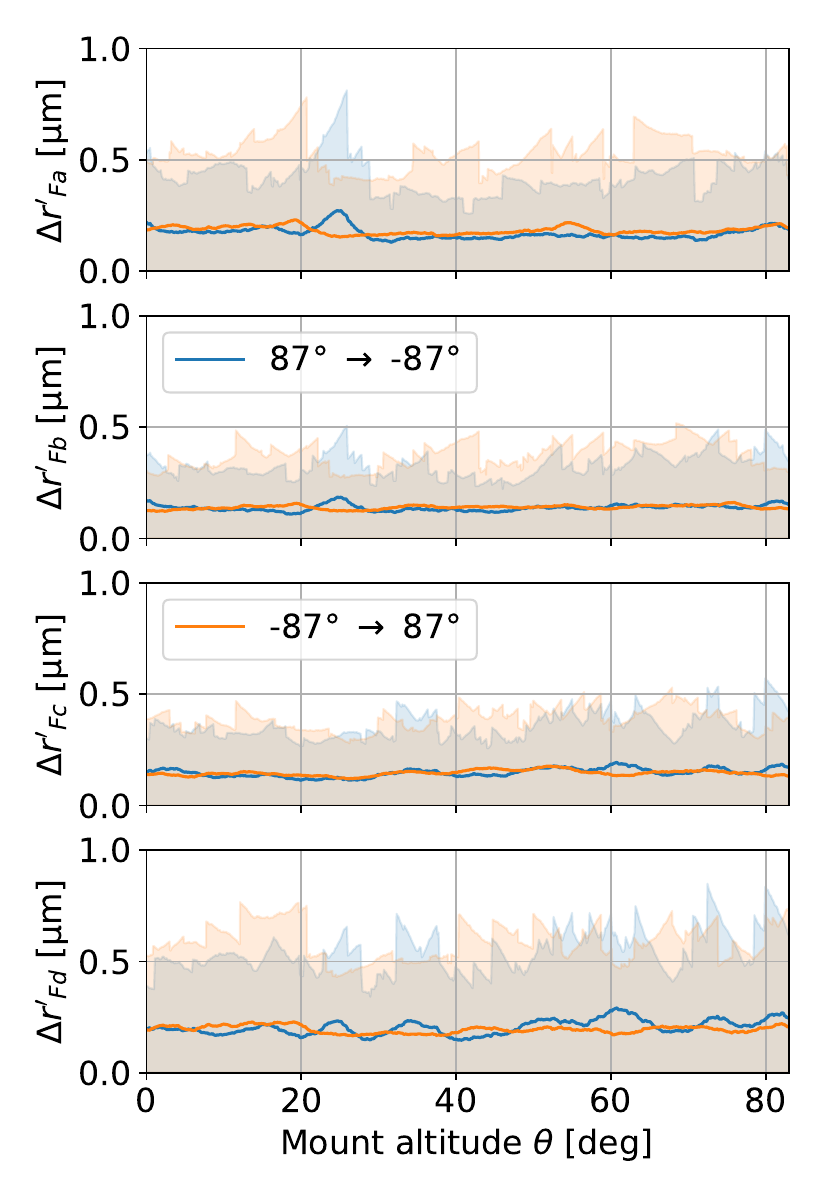}
        \caption{%
          Average (solid lines) and maximum (shaded area) of the absolute deviation within a \SI{7.6}{\degree} tracking window centered around a given $\theta$.
          This windows corresponds to the maximum tracking during a \SI{30}{\minute} MUST exposure.
        }
        \label{fig:calibration_test_MUST}
    \end{subfigure}
    \caption{%
      Investigation of the calibration stability and measurement uncertainty, by using 4 out of 5 fixed spots for calibration, and the 5th spot as a test spot.
      Deviations are calculated according to~\eqref{eq:deviation_full}-\eqref{eq:deviation_max}
    }
    \label{fig:calibration_test}
\end{figure}

In order to evaluate the measurement precision of the setup, and especially its stability at different mount orientations $\theta$, we carry out two mount tracking test measurements.
In the first measurement the mount is set to move at sidereal rate from $\theta = \SIrange{87}{-87}{\degree}$, while in the second measurement the opposite movement $\theta = \SIrange{-87}{87}{\degree}$ is carried out, each movement taking about \SI{12}{\hour}.
Pictures are taken multiple times per minute.
\begin{figure}
    \centering
    \includegraphics[width=\textwidth]{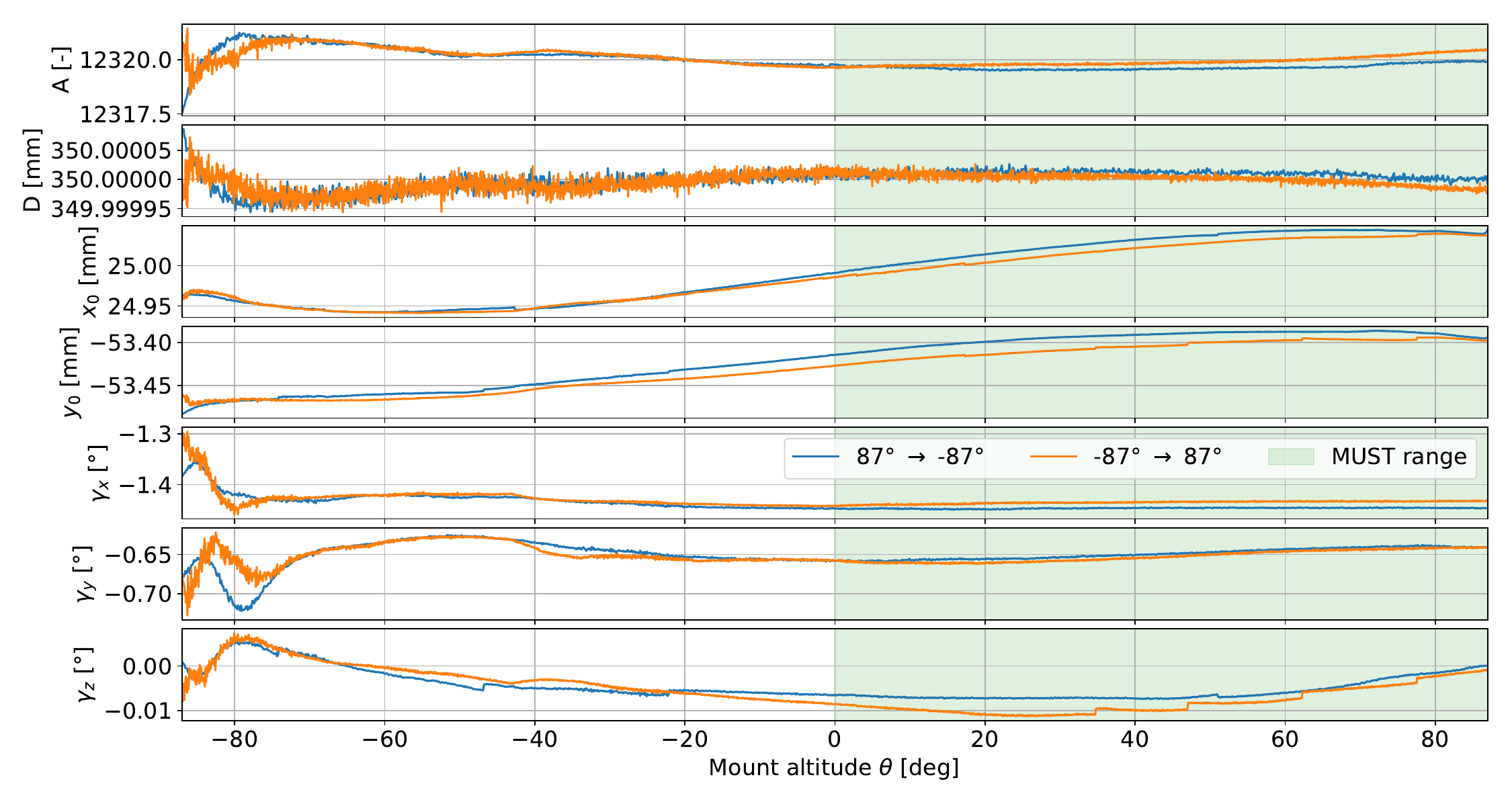}
    \caption{%
      Calibration parameters extracted using all five calibration spots during tracking motions from \SIrange{87}{-87}{\degree} and the reverse motion from \SIrange{-87}{87}{\degree}.
      The source of the discontinuities in the curves is unknown, but possibly attributed to external factors.
      The calibration transform properly absorb these effects, as spot locations after calibration do not show any discontinuities (see for example figure~\ref{fig:calibration_test_Full}).
    }
    \label{fig:calibration_all_parameters}
\end{figure}
The extracted calibration parameters, using all five calibration spots ($D$, $Fa$, $Fb$, $Fc$ and $Fd$) are shown in figure~\ref{fig:calibration_all_parameters}.
All parameters vary smoothly over the entire range, with the forward and backward measurement resulting in nearly the same calibration values.
Some calibration parameters show discrete jumps which are distinct for the forward and backward movement.
The source of these jumps is unknown and could be attributed to external factors, but as the calibration properly accounts for these jumps they are not further studied here.
This can be seen in figure~\ref{fig:calibration_test_Full}, where the relative spot shift after calibration does not show any discontinuities.
All parameters show drastic sudden changes in the range $\theta \in \SIrange{-87}{-70}{\degree}$, independent of the mount movement direction.
A large increase in the variation of the fixed spot positions after calibration is also observed in this region, as can be seen in figure~\ref{fig:calibration_test_Full}.
This indicates that additional non-linear distortions might be present in this elevation range, which cannot be captured by the linear calibration transfer function~\eqref{eq:rotation_calibration}.
The source of this artefact is currently being investigated, but as it lies well outside of the range of interest for MUST (with $\theta_\text{MUST} \in \SIrange{0}{90}{\degree}$) it does not hinder the current measurement campaign.

We use the same two datasets (tracking in both motion directions) to also carry out four partial calibrations, which each use the spot $D$ and three of the four spots $Fa$, $Fb$, $Fc$ and $Fd$.
For each sub-calibration, we treat the spot not used for calibration $S$ (for example $Fa$ for one of them) as a test spot, for which we calculate
\begin{equation}
  \Delta{}r_{S}(\theta) = \left| \vec{r}_{S}(\theta) - \langle\vec{r}_{S}\rangle \right|. \label{eq:deviation_full}
\end{equation}
Figure~\ref{fig:calibration_test_Full} shows the resulting curves for the four test cases.
We observe that over the entire range (excluding \SIrange{-87}{-70}{\degree}) we have a systematic residual of up to \SI{4}{\micro\meter} when measuring a spot which is assumed to be static.
For the tracking stability measurements for MUST (as discussed in section~\ref{sec:intro}), we need to show that the \ac{DUT} fiber spots do not move by more than \SI{\pm1}{\micro\meter} during a maximum observation duration of \SI{30}{\minute}, which represents a maximum altitude change of $\Delta{}\theta_{max} \approx \SI{7.6}{\degree}$.
Accordingly we reprocess the data in figure~\ref{fig:calibration_test} to calculate the stability within $\SI{7.6}{\degree}$ segments
\begin{align}
  \Delta{}r'_{S, max}(\theta) &= \max\left(\left| \vec{r}_{S}(\theta') - \langle\vec{r}_{S}\rangle_{[\theta - \SI{3.8}{\degree},\theta + \SI{3.8}{\degree}]} \right| ~ \text{with} ~ \theta - \SI{3.8}{\degree} \leq \theta' \leq \theta + \SI{3.8}{\degree}\right) \label{eq:deviation_max} \\
  \Delta{}r'_{S, avg}(\theta) &= \avg\left(\left| \vec{r}_{S}(\theta') - \langle\vec{r}_{S}\rangle_{[\theta - \SI{3.8}{\degree},\theta + \SI{3.8}{\degree}]} \right| ~ \text{with} ~ \theta - \SI{3.8}{\degree} \leq \theta' \leq \theta + \SI{3.8}{\degree}\right) \label{eq:deviation_avg}
\end{align}
The resulting curves are shown in figure~\ref{fig:calibration_test_MUST}, where the solid lines represent $\Delta{}r'_{S, avg}(\theta)$  and  the shaded area represents $\Delta{}r'_{S, max}(\theta)$.
We can see that in the relevant $\theta$-range for MUST, the measurement uncertainty from our test setup during any altitude tracking of $\Delta{}\theta_{max}$ reaches a maximum of ca.~\SI{0.5}{\micro\meter}, well below the \SI{1}{\micro\meter} requirement, thus fulfilling the testing requirements.

\section{Results with a 6-Positioner Module from Orbray}
\label{sec:results}

\subsection{XY-stability over the Full MUST Altitude Range}
\label{sec:results_track_full}
We measure the XY-stability of an early prototype module populated with 6 positioners (shown in figure~\ref{fig:positioner_illustration_module}).
This module was fabricated by Orbray Co.,~Ltd. in the context of an Innosuisse Project led by the Astrobots team at EPFL~\cite{galal_prototyping_2025}.
This prototype follows the Trillium design~\cite{silber_25000_2022,silber_design_2022}, with groups of three positioners being integrated together, and the $\Theta$ and $\Phi$ axis of the positioners being coupled.
Accordingly our \ac{DUT} is made up of two Trillium groups.

\begin{figure}
    \centering
    \includegraphics[width=\textwidth]{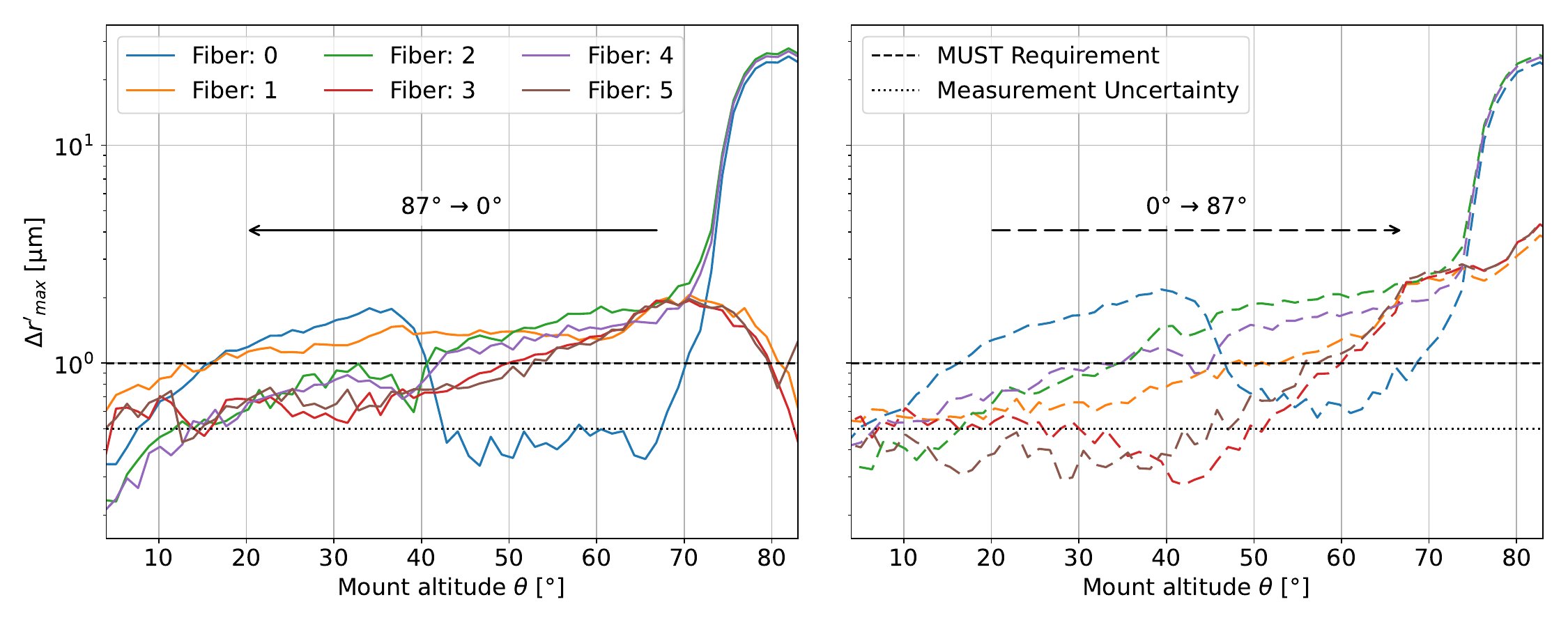}
    \caption{%
      Maximum fiber spot deviation per \SI{7.6}{\degree} tracking window centered on $\theta$ for the Orbray-6 positioners, calculated according to~\eqref{eq:deviation_max}.
      The \ac{MUST} requirement of a stability better than \SI{1}{\micro\meter} is also plotted, as well as the measurement uncertainty estimated from figure~\ref{fig:calibration_test_MUST}.
      All positioners exceed the maximum allowed deviation in some orientations.
    }
    \label{fig:orbray_tracking_full_windows}
\end{figure}

We measured the \ac{DUT} fiber spot stability with the telescope mount tracking at sidereal rate from \SIrange{87}{0}{\degree} and from \SIrange{0}{87}{\degree}.
The robotic positioners were kept static (no $\alpha$/$\beta$ motion) during the entire measurement.
\begin{figure}
    \begin{subfigure}[t]{0.39\textwidth}
        \centering
        \includegraphics[width=\textwidth,clip,trim=0 0 0.5cm 0]{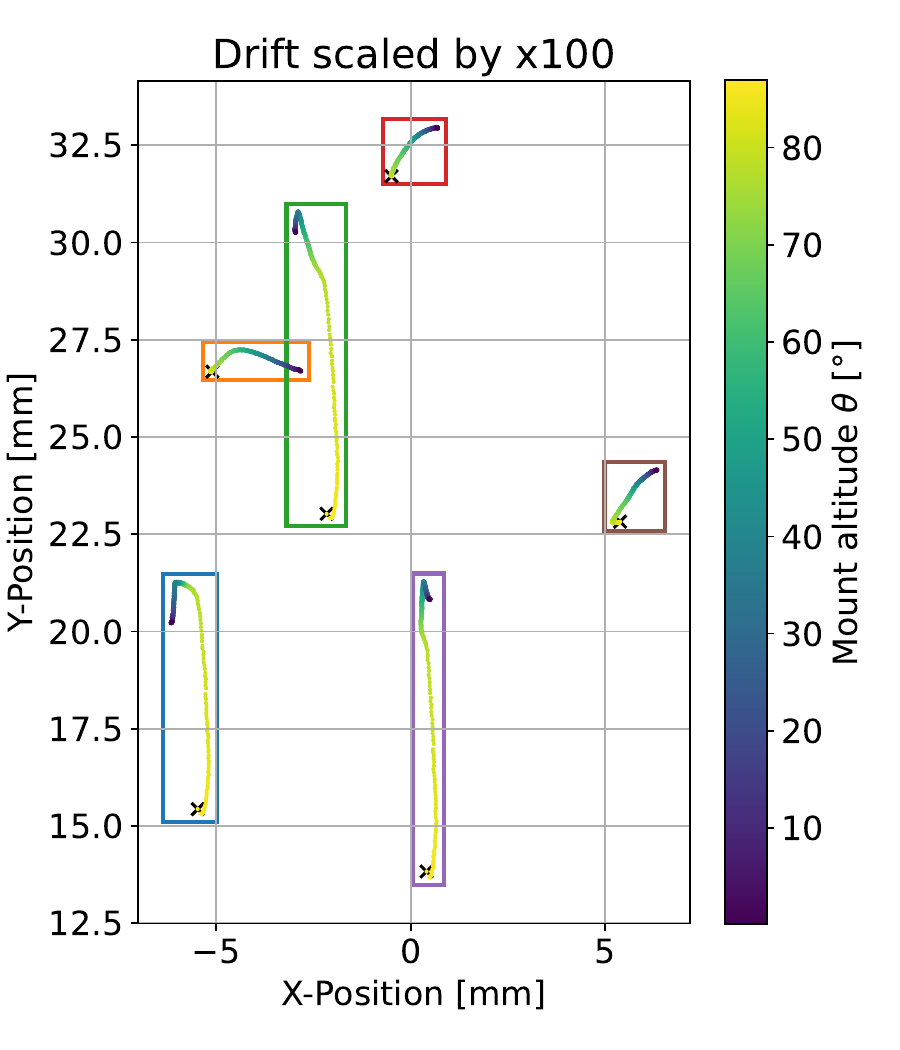}
        \caption{%
          X-Y scatter plot of the fiber spot positions for the measurement \SIrange{87}{0}{\degree}.
          The drift of each fiber within the color outlines is multiplied by a factor of 100.
          The relative position between fibers is only accurate for $\theta = \SI{87}{\degree}$, which is marked with an 'x' for each fiber.
        }
        \label{fig:orbray_tracking_full_XY}
    \end{subfigure}%
    \hfill
    \begin{subfigure}[t]{0.585\textwidth}
        \centering
        \includegraphics[width=\textwidth]{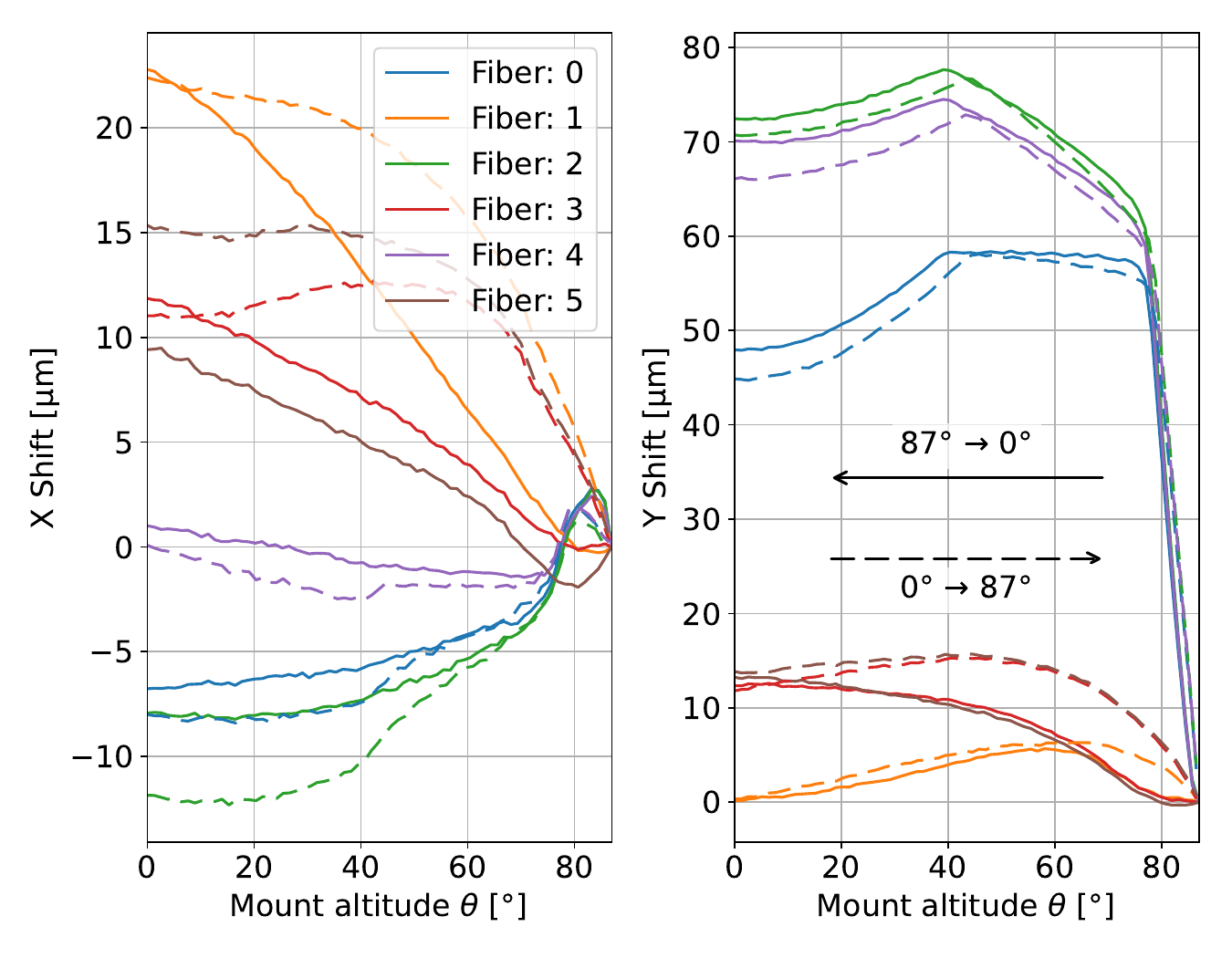}
        \caption{Shift of each fiber spot as a function of $\theta$ relative to the position at $\theta = \SI{87}{\degree}$.
        Data for measurements with increasing and decreasing altitudes are shown, as indicated by the arrows (solid vs. dashed lines).
        Fibers belonging to the same Trillium show correlated motion (Trillium~A: 0, 2, 4 -- Trillium~B: 1, 3, 5).
        }
        \label{fig:orbray_tracking_full_angular}
    \end{subfigure}
    \caption{%
      Results from the full range mount tracking measurements (\SIrange{0}{87}{\degree}) with the Orbray-6 prototype module.
      The colors attributed to fibers are consistent between figures~(a) and~(b).
    }
    \label{fig:orbray_tracking_full}
\end{figure}
Figure~\ref{fig:orbray_tracking_full_XY} shows a scatter plot of the \ac{DUT} fiber positioner during the \SIrange{87}{0}{\degree} measurement.
In order to better visualize the overall movement pattern, the relative drift of each individual fiber spot is amplified by a factor of 100 in this plot, relative to the fiber position at \SI{87}{\degree} (marked by 'x' markers).
We can clearly distinguish two groups of positioners (fibers 0, 2, 4 and fibers 1, 3, 5) each corresponding to one of the two Trillium groups in our module.
In figure~\ref{fig:orbray_tracking_full_angular} the X and Y shift (relative to the position at \SI{87}{\degree}) for both measurements is shown for all fibers.
All fibers exhibit a moderate drift in the X-direction of the order of \SI{10}{\micro\meter} over the \SIrange{87}{0}{\degree} range.
On the other hand, in Y-direction one Trillium shows a large drift up to \SI{80}{\micro\meter}, dominated by a larger initial shift when approaching \SI{87}{\degree}.
The observed fiber spot shifts are to the first order independent of the telescope tracking direction.
Analogous to the analysis in section~\ref{sec:stability} we also calculate here the average and maximum spot deviation for each $\Delta{}\theta = \SI{7.6}{\degree}$ segment, according to \eqref{eq:deviation_avg} and~\eqref{eq:deviation_max}.
The resulting curves are shown in figure~\ref{fig:orbray_tracking_full_windows}.
We can see that all the positioners at some orientations violate the specified MUST limit of \SI{<1}{\micro\meter} of spot deviation over a \SI{30}{\minute} observation (with a maximum altitude change of $\Delta{}\theta = \SI{7.6}{\degree}$).
One set of positioners belonging to the same Trillium structure show drifts of close to \SI{20}{\micro\meter} within windows close to the zenith orientation ($\theta \rightarrow \SI{90}{\degree}$).
These results are independent of the telescope tracking direction.
It should be noted that, as discussed in section~\ref{sec:calibration_outofplane}, our setup can fundamentally not distinguish between a lateral drift of spots along the XY plane, and a large shift along the axial Z direction.
With the large apparent lateral motion observed (in parts exceeding \SI{10}{\micro\meter}), the corresponding relative axial shift would need to be on the order of $\gtrsim \SI{100}{\micro\meter}$ (see table~\ref{tab:dz_summary}) which would by far exceed the defocus requirement of MUST.

\subsection{XY-Stability in Simulated Telescope Operation}
\label{sec:results_track_multi}

\begin{figure}
    \centering
    \includegraphics[width=\textwidth]{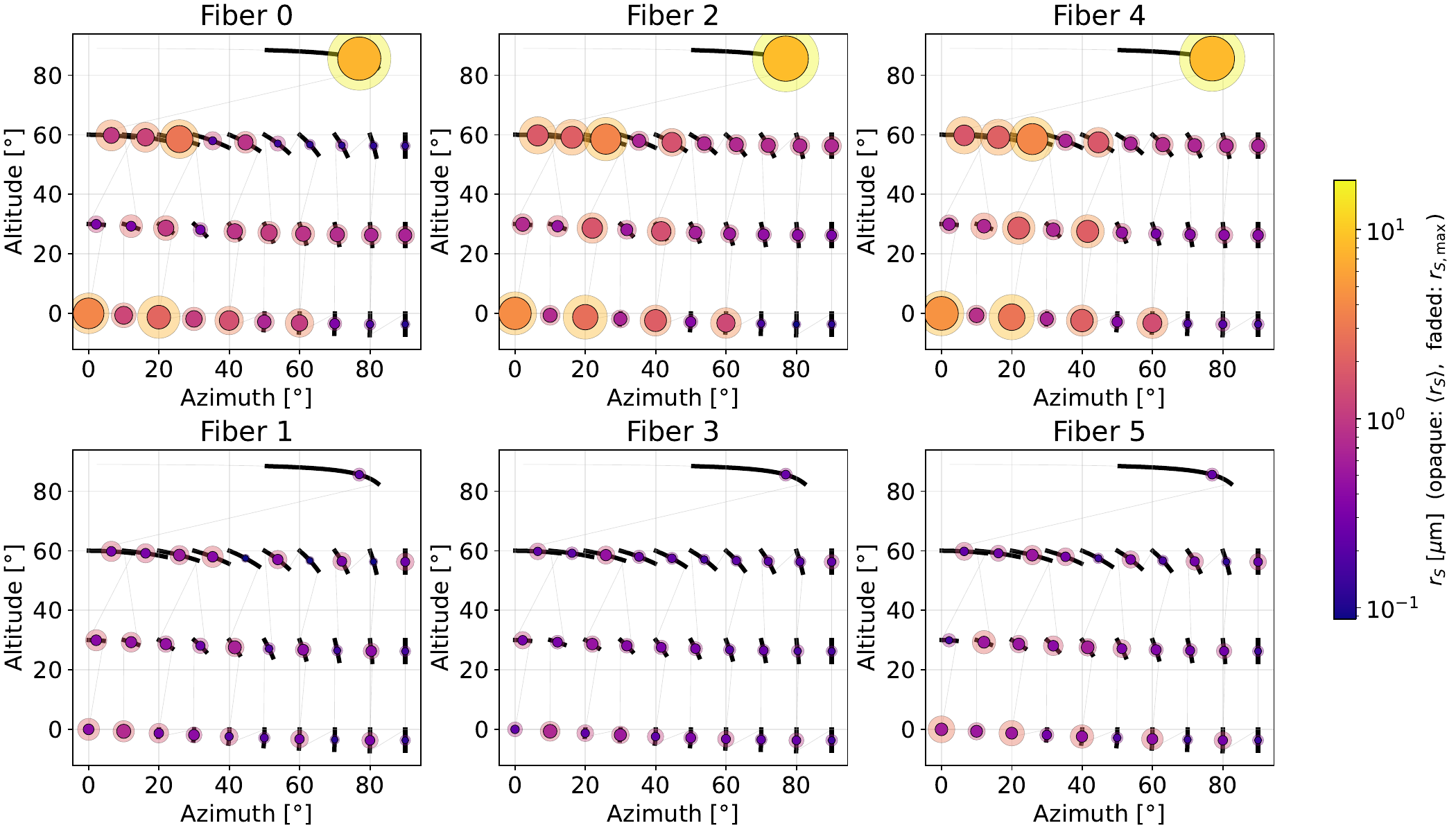}
    \vspace{0.1cm}
    \caption{%
      Fiber spot deviation for the Orbray-6 positioners for each \SI{30}{\minute} tracking session.
      The test orientations are chosen in the \ac{MUST} altitude range with $\SI{0}{\degree}<\theta<\SI{90}{\degree}$.
      The mean radius deviations $\Delta r_S$ are calculated according to~\eqref{eq:deviation_mean_multi} and maximum radius deviations according to~\eqref{eq:deviation_max_multi}.
      Each star tracking segment is represented by a data scatter point at the median $\theta,\phi$ of the segment.
      The scatter point sizes represent the average spot deviations and the shaded circle represents the maximum spot deviation.
      Positioners belonging to the same Trillium are on the same row.
    }
    \label{fig:orbray_multitracking_std_trackrange}
\end{figure}

In realistic survey telescope operation such as for \ac{MUST}, the telescope, including the positioners on the focal plane, slews to a new sky region for each acquisition, followed by an exposure on the order of 10s of minutes.
% MUST-gray 1hour per target; MUST-dark 2.5 hour https://arxiv.org/html/2411.07970v3
We emulate such an operation schedule by iteratively going to a new (altitude $\theta$, azimuth $\phi$) combination and starting a new \SI{30}{\minute} tracking motion at that location, with a focus on the \ac{MUST} altitude range $\SI{0}{\degree}<\theta<\SI{90}{\degree}$.
During each tracking segment $\rm seg_i$ the positions of all fiber spots are recorded, and the average and maximum spot deviation is calculated equivalently to~\eqref{eq:deviation_max} and~\eqref{eq:deviation_avg}:
\begin{align}
  \Delta r_{S,max}^{\rm seg_i}(t) &= \max\left(\left|\svec{r}{\rm seg_i}{S}(t) - \langle\svec{r}{\rm seg_i}{S}\rangle_{\SI{30}{min}}\right| \right).\label{eq:deviation_max_multi} \\
  \Delta r_{S,avg}^{\rm seg_i}(t) &= \avg\left(\left|\svec{r}{\rm seg_i}{S}(t) - \langle\svec{r}{\rm seg_i}{S}\rangle_{\SI{30}{min}}\right| \right).\label{eq:deviation_mean_multi}
\end{align}
Figure~\ref{fig:orbray_multitracking_std_trackrange} shows the tracking motion range in $(\theta,\phi)$ space, and the associated average and maximum fiber drift for each fiber in each tracking segment.
\begin{figure}
    \centering
    \includegraphics[width=\textwidth,clip,trim=0 2cm 0 2cm]{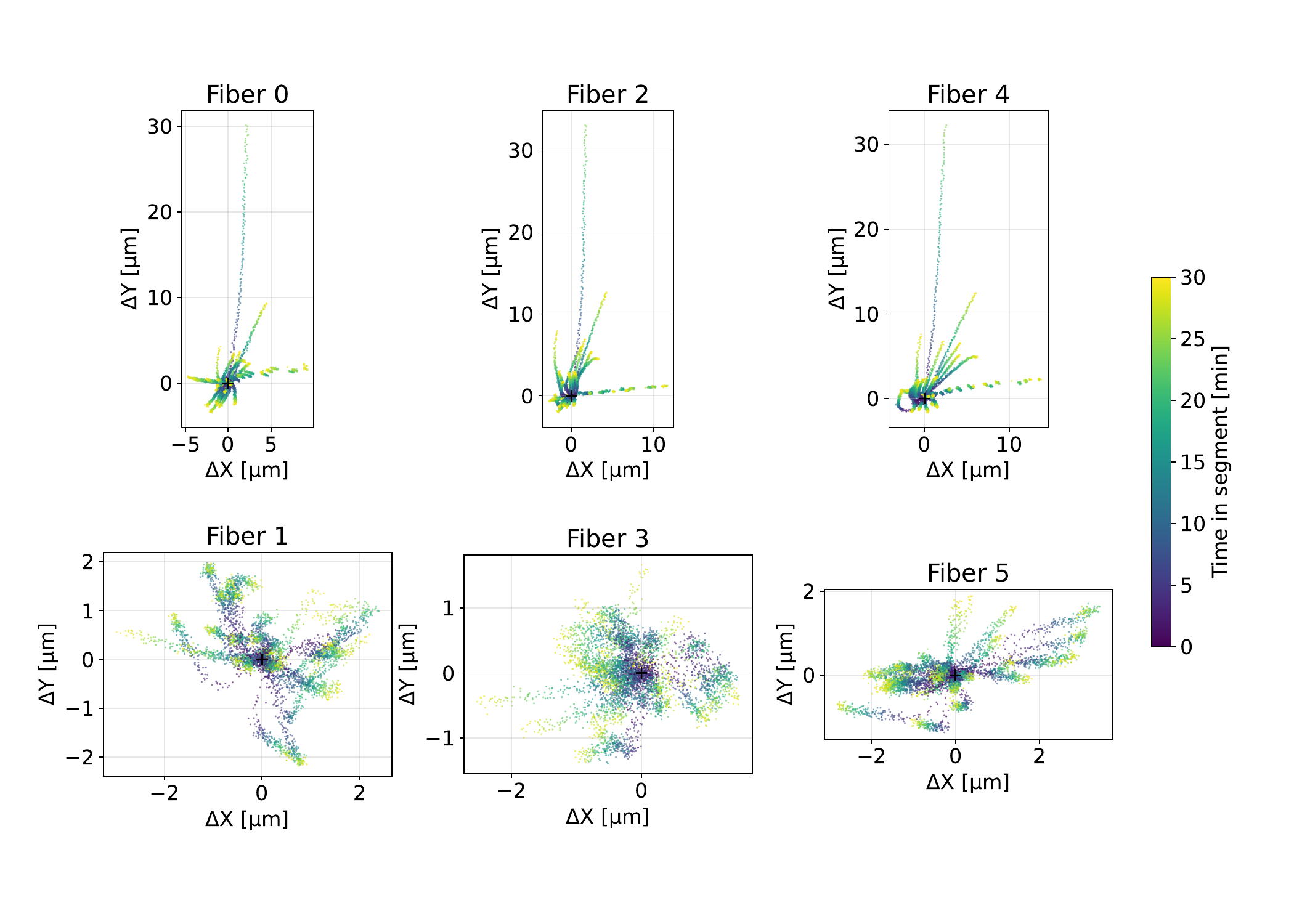}
    \caption{%
      X-Y scatter plot of the fiber spot positions for all star-tracking segments stacked, relative to the starting position $\svec{r}{\rm seg_i}{S}(0)$ of each segment.
      The tracking time is \SI{30}{\minute} for each segment, and color bar indicates the elapsed time during each segment.
      Positioners belonging to the same Trillium are on the same row.
    }
    \label{fig:orbray_multitracking_drift_overlay}
\end{figure}
Here we observe again a correlated deviation pattern among positioners belonging to the same Trillium raft (fibers~0,~2,~4 vs. fibers~1,~3,~5), with fibers 0,~2,~4 showing significantly higher drift as previously observed.
The tracking segment with mount orientation $\theta \approx \SI{90}{\degree}$ shows a very high deviation in fibers 0,~2,~4, which is consistent with the observation of initial large motion around $\theta = \SI{87}{\degree}$ in the full tracking range, as discussed in section~\ref{sec:results_track_full}.
We can see this feature more clearly in figure~\ref{fig:orbray_multitracking_drift_overlay}, where positioner drifts from all tracking segments are overlaid, with $\Delta{}x/\Delta{}y$ relative to the position $\svec{r}{\rm seg_i}{S}(0)$ at the start of each segment.
The large drifts in the Y direction from one segment are consistent with the XY drift from the full range tracking test in figure~\ref{fig:orbray_tracking_full_XY}.
Relatively large deviations captured in fibers 0,~2,~4 also occur at mount orientation $\theta= \phi = \SI{0}{\degree}$ (near telescope home position), where the tracking motion translates primarily into a rotation around the optical axis.
This is an important addition to the full range tracking test, which fixed the azimuth angle to $\phi = \SI{90}{\degree}$, indicating that the positioners have a non-trivial response to a self-rotation motion, which is common to the star tracking operation of survey telescopes.
One important note is that for this measurement the equatorial mount was set to latitude \SI{0}{\degree}, corresponding to a telescope placed at the equator.
For future measurements this will be adjusted to faithfully represent the \ac{MUST} site.

\section{Conclusions}
\label{sec:conclusions}
Lessons learned from previous spectroscopic surveys have shown that testing positioners in conditions similar to what they would experience on the telescope is essential to guarantee successful long-term operations.
We present a test setup for orientation dependent testing of robotic positioner modules for future Stage-V spectroscopic surveys.
Focusing on XY measurements, we discussed the general setup of our test stand, and the identification and mitigation of various effects which can degrade the measurement result, such as image stacking for shot noise reduction and the installation of PMMA screens to reduce thermal turbulence inside our dark box.
A dedicate calibration routine is implemented, making use of five fixed spots to correct for any thermal or mechanical deformation of the test stand.
We demonstrate our measurement precision from cross-calibration of fixed fiducial spots, and found a residual uncertainty of ca.~\SI{0.5}{\micro\meter}, during a stability measurement while the telescope mount is tracking at sidereal rate over $\Delta{}\theta = \SI{7.6}{\degree}$.
Finally we presented XY-stability results for an early 6-positioner prototype module fabricated by Orbray Co., Ltd.
The results show that all 6 positioners exceed the required maximum relative position drift of \SI{1}{\micro\meter} for $\Delta{}\theta = \SI{7.6}{\degree}$, especially at mount orientations close to $\theta = \SI{90}{\degree}$ where the positioners point nearly straight upwards.
We further emulated realistic survey operation through a sequence of \SI{30}{\minute} star-tracking segments distributed across the \ac{MUST} altitude and azimuth range.
The results are overall consistent with the full-range tracking test, showing the same orientation-dependent, Trillium-correlated drift and likewise exceeding the \SI{1}{\micro\meter} limit.
These findings emphasize the importance of early and systematic testing of positioner prototypes for future spectroscopic surveys, especially regarding the effect of changing gravity vectors.
We have shown that our test stand is ready to contribute to these efforts, and we are working on expanding our orientation dependent testing capabilities towards defocus and tilt testing, as well as carrying long term reliability tests under constantly changing gravity directions.

\appendix

\acknowledgments
We acknowledge the Astrobots team of the \ac{EPFL} for providing us with the Orbray-6 prototype module.
We also thank them for the many fruitful discussions.
This work is supported by the Swiss National Science Foundation FLARE grant no. 232756 and the University of Zurich.

% References
\bibliography{20260504_SPIEGeneralPaper} % bibliography data in report.bib
\bibliographystyle{spieref} % makes bibtex use spiebib.bst

\end{document}